\documentclass[runningheads]{llncs}

\usepackage{eccv}

\usepackage{eccvabbrv}

\usepackage{graphicx}
\usepackage{booktabs}

\usepackage{multirow}
\usepackage{fancyhdr}

\newcommand{\eone}{\mathbf{E^1}}
\newcommand{\etwo}{\mathbf{E^2}}
\newcommand{\ethree}{\mathbf{E^3}}
\newcommand{\efour}{\mathbf{E^4}}
\newcommand{\znoise}{\mathbf{z_{noise}}}

\usepackage[accsupp]{axessibility}  

\usepackage{hyperref}

\usepackage{orcidlink}

\begin{document}

\title{Hierarchical Conditioning of Diffusion Models Using Tree-of-Life for Studying Species Evolution} 

\titlerunning{Hierarchically Conditioned Diffusion Model}

\author{Mridul Khurana\inst{1}\orcidlink{0009-0003-9346-3206} \and
Arka Daw\inst{2}\orcidlink{0009-0006-3319-1271} \and 
M. Maruf \inst{1} \orcidlink{0000-0003-0637-1753} \and 
Josef C. Uyeda \inst{1} \orcidlink{0000-0003-4624-9680} \and 
Wasila Dahdul \inst{3}\orcidlink{0000-0003-3162-7490} \and 
Caleb Charpentier \inst{1} \orcidlink{0000-0002-9787-7081} \and 
Yasin Bakış \inst{4}\orcidlink{0000-0001-6144-9440} \and 
Henry L. Bart Jr. \inst{4}\orcidlink{0000-0002-5662-9444} \and 
Paula M. Mabee \inst{5}\orcidlink{0000-0002-8455-3213} \and 
Hilmar Lapp \inst{6}\orcidlink{0000-0001-9107-0714} \and 
James P. Balhoff \inst{7}\orcidlink{0000-0002-8688-6599} \and 
Wei-Lun Chao \inst{8}\orcidlink{0000-0003-1269-7231} \and 
Charles Stewart \inst{9}\orcidlink{0000-0001-6532-6675} \and 
Tanya Berger-Wolf\inst{8}\orcidlink{0000-0001-7610-1412} \and 
Anuj Karpatne \inst{1}\orcidlink{0000-0003-1647-3534}
}

\authorrunning{M. Khurana et al.}


\institute{ Virginia Tech, Blacksburg VA, USA \\
            \email{\{mridul, karpatne\}@vt.edu} \and 
            Oak Ridge National Laboratory, Oak Ridge TN, USA \and
            University of California, Irvine CA, USA \and
            Tulane University, New Orleans LA, USA \and
            Battelle, Columbus OH, USA \and
            Duke University, Durham NC, USA \and
            University of North Carolina at Chapel Hill, Chapel Hill NC, USA \and
            The Ohio State University, Columbus OH, USA \and
            Rensselaer Polytechnic Institute, Troy NY, USA 
}

\maketitle

\pagestyle{fancy}
\fancyhf{} 
\fancyhead[L]{Published as a conference paper at ECCV 2024}
\fancyfoot[C]{\thepage}

\setcounter{footnote}{0}

\begin{abstract}
A central problem in biology is to understand how organisms evolve and adapt to their environment by acquiring variations in the observable characteristics or traits of species across the tree of life. 
With the growing availability of large-scale image repositories in biology and recent advances in generative modeling, there is an opportunity to accelerate the discovery of evolutionary traits automatically from images. 
Toward this goal, we introduce Phylo-Diffusion, a novel framework for conditioning diffusion models with phylogenetic knowledge represented in the form of HIERarchical Embeddings (HIER-Embeds). 
We also propose two new experiments for perturbing the embedding space of Phylo-Diffusion:  trait masking and trait swapping, inspired by counterpart experiments of gene knockout and gene editing/swapping.
Our work represents a novel methodological advance in generative modeling to structure the embedding space of diffusion models using tree-based knowledge. Our work also opens a new chapter of research in evolutionary biology by using generative models to visualize evolutionary changes directly from images.
We empirically demonstrate the usefulness of Phylo-Diffusion in capturing meaningful trait variations for fishes and birds, revealing novel insights about the biological mechanisms of their evolution. 
\footnote{Model and code can be found at \href{https://imageomics.github.io/phylo-diffusion/}{\texttt{imageomics.github.io/phylo-diffusion}}}
  
\end{abstract}

\section{Introduction}
\label{sec:intro}

\begin{figure}
\centering
\includegraphics[width=1\linewidth]{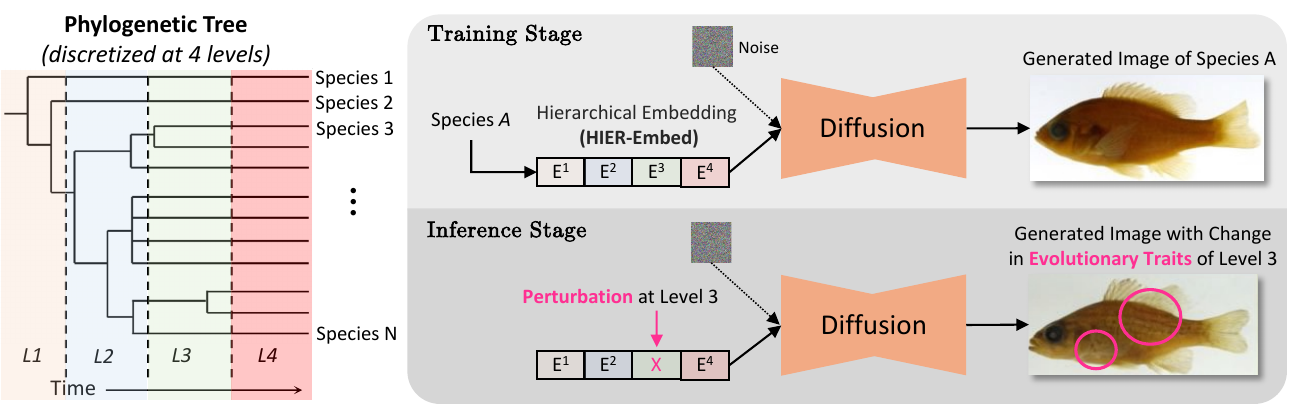}
\caption{Overview of Phylo-Diffusion framework. Every species in the tree of life (phylogenetic tree) is encoded to a HIERarchical Embedding (HIER-Embed) comprising of four vectors (one for each phylogenetic level), which is used to condition a latent diffusion model to generate synthetic images of the species. By structuring the embedding space with phylogenetic knowledge, Phylo-Diffusion enables visualization of changes in the evolutionary traits of a species (circled pink) upon perturbing its embedding.}
\label{fig:architechture}
\end{figure}

Given the astonishing diversity of life forms on the planet, an important end goal in biology is to understand how organisms evolve and adapt to their environment by acquiring variations in their observable characteristics or \textit{traits} (\eg, {beak color, stripe pattern, and fin curvature}) over millions of years in the process of evolution. Our knowledge of species evolution is commonly represented in a graphical form as the ``tree of life'' (also referred to as the \textit{phylogenetic tree} \cite{Kapli2020}, see Figure \ref{fig:architechture}), illustrating the evolutionary history of species (leaf nodes) and their common ancestors (internal nodes). Discovering traits that are heritable across the tree of life, termed \textit{evolutionary traits}, is important for a variety of biological tasks such as tracing the evolutionary timing of trait variations common to a group of species and analyzing their genetic underpinnings through gene-knockout or gene-editing/swapping (\eg, CRISPR \cite{crispr}) experiments. However, quantifying trait variations across large groups of species is labor-intensive and time-consuming, as it relies on expert visual attention and subjective definitions \cite{simoes2017giant}, hindering rapid scientific advancement \cite{lurig2021}. 

The growing deluge of large-scale image repositories in biology \cite{inat2021, stevens2024bioclip, gharaee2024step} presents a unique opportunity for machine learning (ML) methods to accelerate the discovery of evolutionary traits automatically from images. In particular,  with recent developments in generative modeling such as latent diffusion models (LDMs) \cite{rombach2022high}, we are witnessing rapid improvements in our ability to control the generation of high-quality images based on input conditioning of text or image prompts. This is facilitating breakthroughs in a variety of commercial use-cases of computer vision where we can analyze how changes in the input prompts affect variations in the generated images \cite{ruiz2023dreambooth, zhang2023adding, esser2024scaling}.
We ask the question: \emph{can we leverage LDMs to control the generation of biological images of organisms conditioned on the position of a species in the tree of life?} In other words, can we encode the structure of evolutionary relationships among species and their ancestors as input conditions in LDMs?
This can help us analyze trait variations in generated images across different branches in the phylogenetic tree, revealing novel insights into the biological mechanisms of species evolution.

Toward this goal, we introduce \textbf{Phylo-Diffusion}, a novel framework for discovering evolutionary traits of species from images by conditioning diffusion models with phylogenetic knowledge (see Figure \ref{fig:architechture}). One of the core innovations of Phylo-Diffusion is a novel HIERarchical Embedding (\textbf{HIER-Embed}) strategy that encodes evolutionary information of every species as a sequence of four vectors, one for each discretized level of ancestry in the tree of life (covering different evolutionary periods). 
We also propose two novel experiments for analyzing evolutionary traits by perturbing the embedding space of Phylo-Diffusion and observing changes in the features of generated images, akin to biological experiments involving genetic perturbations.
First, we introduce \textbf{Trait Masking}, where one or more levels of information in HIER-Embed are masked out with noise to study the disappearance of traits inherited by species at those levels. This is inspired by  \textit{gene knockout} experiments \cite{griffiths2005introduction}, wherein one or more genes are deactivated or ``knocked out'' to investigate the gene's function, particularly its impact on the traits of the organism. Second, we introduce \textbf{Trait Swapping}, where a certain level of HIER-Embed in a reference species is swapped with the embedding of a sibling node at the same level, similar in spirit to \textit{gene editing/swapping} experiments made possible by the CRISPR technology \cite{crispr}. The goal of trait swapping is to visualize trait differences at every branching point in the tree of life that results in the diversification of species during evolution.

Here are the main contributions of this paper. Our work represents a novel methodological advance in the emerging field of knowledge-guided machine learning (KGML) \cite{karpatne2024knowledge,karpatne2022knowledge,karpatne2017theory} to structure the embedding space of generative models using tree-based knowledge. Our work also opens a new chapter of research in evolutionary biology by using generative models to visualize evolutionary changes directly from images, which can serve a variety of biological use-cases. For example, Phylo-Diffusion can help biologists automate the discovery of \textit{synapomorphies}, which are distinctive traits that emerge on specific evolutionary branches and are crucial for systematics and classification \cite{zelditch1995morphometrics}. Our proposed experiments of trait masking and swapping can also be viewed as novel image-based counterparts to genetic experiments, which traditionally take years. Our work thus enables biologists to rapidly analyze the impacts of genetic perturbations on particular branches of the phylogenetic tree--a grand challenge in developmental biology \cite{edmunds2015phenoscape,manda2015using}.
We empirically demonstrate the usefulness of Phylo-Diffusion in capturing meaningful trait changes upon perturbing its embedding for fishes and birds, generating novel hypotheses of their evolution.
\section{Related Works and Background}

 \paragraph{\textbf{Interpretable ML:}}
Discovering evolutionary traits from images requires the identification and interpretation of fine-grained features in images that define and differentiate species. Several methodologies have recently been developed in the field of interpretable ML for localizing image regions that contain discriminatory information of classes, including ProtoPNet \cite{chen2019looks}, PIP-Net \cite{Nauta_2023_CVPR} and INTR \cite{paul2023simple}. 
Despite their effectiveness and applicability across a wide range of applications, these methods are not directly suited for our target application of discovering evolutionary traits for two primary reasons. First, they are not designed to incorporate structured biological knowledge (\eg, knowledge of the tree of life) in the learning of interpretable features, and thus are unable to provide biologically meaningful explanations of feature differences across groups of species in the phylogenetic tree, which is key to discovering evolutionary traits. Second, since most methods in interpretable ML are designed for the task of classification, it is non-trivial to integrate them in generative modeling frameworks to produce synthetic images with controlled perturbations in the embedding space similar to gene knockout and gene editing/swapping experiments, in contrast to our proposed experiments in Phylo-Diffusion.

\paragraph{\textbf{Phylogeny-guided Neural Networks (Phylo-NN):}}

A recent work closely aligned with our goal of discovering evolutionary traits directly from images is Phylo-NN \cite{elhamod2023discovering}. Phylo-NN uses an encoder-decoder architecture to represent images of organisms as structured sequences of feature vectors termed ``Imageomes'', where different segments of Imageomes capture evolutionary information from varying levels of ancestry in the phylogenetic tree. While Phylo-NN shares several similarities with our proposed framework, Phylo-Diffusion, in terms of motivations and problem formulations, there are also prominent differences. The primary goal of Phylo-NN is specimen-level image reconstruction, whereas Phylo-Diffusion considers a different goal of controlling image generation at the species-level. As a result, Phylo-NN learns a unique Imageome sequence for every organism, enabling us to study the variability in individuals from the same species and the analysis of similarity in Imageome segments learned at shared ancestry levels. On the other hand, Phylo-Diffusion learns a unique embedding for every species and ancestor node in the tree of life, which serves as input conditions to generate distributions of synthetic images. Phylo-Diffusion thus uses hard constraints to ensure that all species with a common ancestor learn the exact same embeddings at their shared ancestry levels, making it easy to analyze trait commonalities and variations across groups of species, in contrast to Phylo-NN. 
Additionally, Phylo-Diffusion allows for perturbations in the embedding space of generative models in biologically meaningful ways inspired by gene knockout and gene editing/swapping experiments, going beyond the capabilities of Phylo-NN. We consider Phylo-NN as a baseline in our experiments to compare its performance with Phylo-Diffusion.

\paragraph{\textbf{Background on Latent Diffusion Models (LDMs):}}
One of the state-of-the-art approaches in generative modeling is the framework of Diffusion Models \cite{ho2020denoising}, which learns a target distribution $p(x)$ by incrementally transforming a noisy sample $x$ generated from a Gaussian distribution $\mathcal{N}(0, I)$ into one that is more likely to be generated from $p(x)$ over a series of timesteps $T$.
While early frameworks of diffusion models (\eg, DDPM \cite{ho2020denoising}, DDIM \cite{song2020denoising} and ADM \cite{dhariwal2021diffusion}) suffered from high computational costs and long training/inference times, Latent Diffusion Models (LDMs) \cite{rombach2022high} are able to address these concerns to a large extent by operating in a compressed latent space, significantly accelerating their ability to generate high-resolution images. The basic idea of LDMs is to train a separate auto-encoder to map an input image $x$ into its latent representation $z_0 = \mathcal{E}(x)$ using encoder $\mathcal{E}$, which when fed to decoder $\mathcal{D}$ produces a reconstruction of the original image,  $\tilde{x} = \mathcal{D}(\tilde{z}_0)$. LDMs employ diffusion models in the compressed latent space $z$ by modeling the conditional probability of the reverse diffusion process as $\tilde{z}_{t-1} \sim p_{\theta}(\tilde{z}_{t-1} | \tilde{z}_{t}, y, t)$, where $y$ is the input condition. This is implemented using a conditional denoising U-Net backbone $\epsilon_{\theta}(z_t, y, t)$ with learnable parameters $\theta$.
LDMs also pre-process $y$ using a domain-specific encoder $\mathbf{E} = \tau_{\phi}(y)$ trained alongside the U-Net backbone $\epsilon_{\theta}$ to project $y$ into the intermediate layers of $\epsilon_{\theta}$ using cross-attention mechanisms. The learnable parameters of LDMs are trained by minimizing the following loss function:
\begin{equation}
\mathcal{L}(\theta,\phi) = \mathbb{E}_{z_t, y, t, \epsilon \sim \mathcal{N}(0, I)}\left[ \|\hat{\epsilon}_{\theta}(z_t, \tau_{\phi}(y), t) - \epsilon\|^2 \right]
\label{eq:cond_ldm}
\end{equation}
\section{Proposed Framework of Phylo-Diffusion}

\subsection{Hierarchical Embedding (HIER-Embed)} 
\label{sec:hier-embed}

Phylo-Diffusion uses a novel hierarchical embedding (HIER-Embed) strategy to structure the embedding $E$ of every species node using phylogenetic knowledge. 
As a first step, we consider a discretized version of the phylogenetic tree involving four ancestral levels, level-1 to level-4, where every level corresponds to a different range of time in the process of evolution. (See \Cref{app:phylogy} for a detailed characterization of the four ancestry levels for fish species used in this study.) Given a set of $n$ species,
$\mathcal{S} = \{S_1, S_2, S_3, ..., S_n\}$, let us represent the position of species $S_i \in \mathcal{S}$ in the phylogenetic tree at the four ancestry levels as  $\{S_i^1, S_i^2, S_i^3, S_i^4\}$, where $S_i^l$ represents the ancestor node of $S_i$ at level-$l$. Hence, if two species $S_i$ and $S_j$ share common ancestors till level-$k$, then $S_i^l = S_j^l$ for $l=1$ to $k$.
We define the level-$l$ embedding of species $S_i$ as:
\begin{equation}
\mathbf{E}_{i}^l = \texttt{Embed}(S_i^l)   \in \mathbb{R}^{d'},
\end{equation}
where $\texttt{Embed}(.)$ is a learnable embedding layer that provides a simple way to store and look-up the trained embeddings of every node. The combined hierarchical embedding (HIER-Embed) of species $S_i$ is obtained by concatenating its embeddings across all four levels as follows:

\begin{equation}
    \mathbf{E}_i = \tau (S_i) = \texttt{Concat[} ~\mathbf{E}_{i}^1, ~\mathbf{E}_{i}^2, ~\mathbf{E}_{i}^3, ~\mathbf{E}_{i}^4 ~ \texttt{]} \in \mathbb{R}^d,
\end{equation}
where \texttt{Concat[.]} denotes the concatenation operation and $y=S_i$ is the input condition used in LDMs. Note that different segments of $\mathbf{E}_{i}$ capture information about the traits of $S_i$ acquired at different time periods of evolution. In particular, we expect the embedding vectors learned at earlier ancestry levels of $\mathbf{E}_i$ to capture evolutionary traits of $S_i$ common to a broader group of species. On the other hand, embeddings learned at later ancestry levels are expected to be more specific to $S_i$. 
In the following, we present two novel experiments for studying evolutionary traits by perturbing the embedding space learned by HIER-Embed.

\begin{figure}[t]
\centering
\begin{subfigure}{0.37\linewidth}
\includegraphics[width=\textwidth]{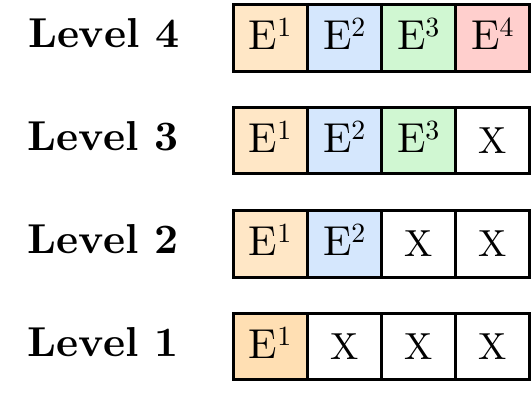}
\caption{Trait Masking}
\label{fig:masking}
\end{subfigure}
\hspace{30pt}
\begin{subfigure}{0.4\linewidth}
\includegraphics[width=\textwidth]{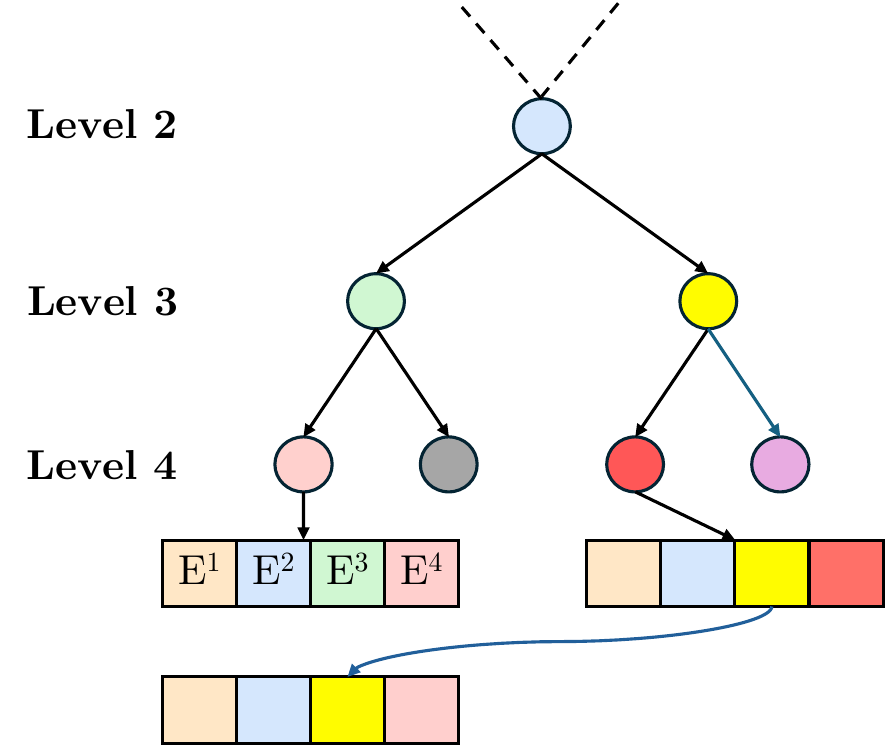}
\caption{Trait Swapping}
\label{fig:swapping}
\end{subfigure}
\caption{Schematic diagrams of the two  proposed experiments for discovering evolutionary traits using Phylo-Diffusion.}
\label{fig:trait_masking_swapping}
\end{figure}

\subsection{Proposed Experiment of Trait Masking}
\label{sec:trait-masking}

The goal of this experiment is to verify if HIER-Embed is indeed able to capture hierarchical information in its level-embeddings such that masking information at lower levels of the embedding only erases traits acquired at later stages of evolution while retaining trait variations learned at earlier levels. In other words, we want to verify that the embeddings learned by HIER-Embed at level-$l$ capture information common to all descendant species that are part of the same sub-tree at level-$l$. 
\Cref{fig:masking} represents a schematic diagram of the process followed for trait masking. We start with the combined embedding containing information at all four levels, $[\eone, \etwo, \ethree, \efour]$. To examine what is learned at the last level of this embedding, we mask it out by substituting it with Gaussian noise defined as $\mathbf{z_{noise}} \sim \mathcal{N}(0, I) \in \mathbb{R}^{d'}$. This results in the perturbed embedding $[\eone, \etwo, \ethree, \znoise]$, effectively eliminating the species-level (or $\efour$) information. This masking should prompt the model to generate images that reflect only the information learned up to the third level while obscuring species-level details. We can extend this experiment by incrementally introducing $noise$ at later levels, \eg, at both levels 3 \& 4, and so on.

\paragraph{\textbf{Expected Changes in Probability Distributions:}} Note that when all four level embeddings are used, \ie $[\eone, \etwo, \ethree, \efour]$, the generated images are expected to be classified to a unique species $S_i$. In terms of probability distributions, the probability of predicting species $S_i$ should be distinctly higher than the probability of predicting any other species. However, when we mask out certain level embeddings (\ie, mask out information at level 4), we are intentionally removing information necessary to distinguish species $S_i$ from its siblings species that are part of the same sub-tree (e.g., those that share a common ancestor at level 3). For this reason, we expect the generated images to show higher probabilities of being classified as any of the descendant species of the sub-tree, compared to the other species that are outside of the sub-tree. 
To quantify this behavior, we can measure the
change in probability distributions for species within and outside the sub-tree after masking out an internal node. Since we expect probabilities to increase only for species within the sub-tree, the mean increase in probabilities for within-subtree species should be higher than that for out-of-subtree species, as empirically demonstrated later in the Results Section.

\subsection{Proposed Experiment of Trait Swapping}
\label{sec:trait-swapping}

In trait swapping, we substitute the level-$l$ embedding of a source species with the level-$l$ embedding of a sibling subtree at an equivalent level. 
\Cref{fig:swapping} shows a schematic representation of the trait swapping experiment 
where the level-3 embedding (green) of a source species is replaced with its sibling level-3 embedding (yellow). 
Images generated for this perturbed embedding are expected to retain all of the traits of the source species except the swapped embedding, which should borrow traits from the sub-tree rooted at the sibling node. Visualizing trait differences in the generated images before and after trait swapping can help us understand the evolutionary traits that branched at a certain level (\eg, those leading to the diversification of green and yellow sub-trees at level-3 in the example phylogeny of Figure \ref{fig:swapping}). 
In terms of the probability distribution, similar to trait masking, we expect to see a drop in the probabilities of the source species (pink), and simultaneously, we expect an increase in probabilities for all the descendent species in subtree at node yellow, \ie red and purple.

\begin{table*}[t]
\begin{tabular}{p{3.4cm}p{3cm}p{3cm}p{1.2cm}}
\toprule
\textbf{Ground Truth} & \textbf{Class \newline Conditional} & \textbf{Scientific \newline Name} & \textbf{Phylo- \newline Diffusion} \\ \midrule
\multicolumn{4}{l}{\includegraphics[width=\textwidth, height=40mm]{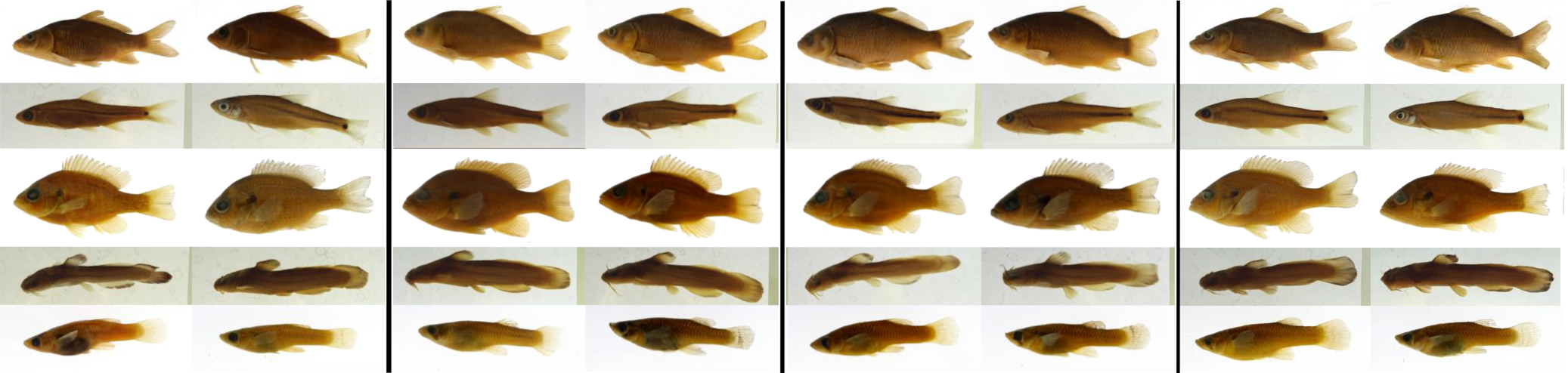}} \\ \bottomrule
\end{tabular}
\captionof{figure}{Comparing the quality of synthetic images generated by different conditioning mechanisms in LDMs. Every row corresponds to a different species and we show two samples per species for every conditioning mechanism. The order of species from top to bottom is \textit{Cyprinus carpio}, \textit{Notropis hudsonius}, \textit{Lepomis auritus}, \textit{Noturus exilis}, and \textit{Gambusia affinis}.}
\label{fig:samples}
\end{table*}

\section{Evaluation Setup}

\paragraph{\textbf{Datasets:}}
We use a collection of fish images as our primary dataset for evaluation. This dataset was procured from the Great Lakes Invasives Network (GLIN) \cite{glin}  Project, comprising a total of 5434 images spanning 38 fish species. We obtained the phylogenetic tree of fish species from \textit{opentree} \cite{10.3897/BDJ.5.e12581} python package (see \Cref{app:phylogy} for details on the phylogenetic tree). The raw museum images were pre-processed and resized to $256 \times 256$ pixels and the dataset was partitioned into training and validation sets, following a 75-25 split. We provide additional results on the CUB-200-2011 dataset \cite{WahCUB_200_2011} of bird species in \Cref{app:cub_results}.

\paragraph{\textbf{Baselines of Conditioning Mechanisms:}}
(1) {\textit{Class Conditional:}} One of the simplest ways of encoding information about a species class is to map class labels $y \in [1, N_c]$ to a fixed $d$-dimensional embedding vector $e \in \mathbb{R}^d$ using a trainable embedding layer. Note that the resulting embeddings are not designed to contain any hierarchical information in contrast to HIER-Embed. (2) {\textit{Scientific Name Encoding:}}  The scientific name of a species contains valuable biological information typically comprising of a combination of the \emph{genus} name and \emph{species} name. Since species that share their \emph{genus} name are likely to contain common phylogenetic traits, we use them as a baseline for conditioning LDMs for discovering evolutionary traits. Specifically, we employ a pre-trained frozen CLIP model \cite{radford2021learning} to encode the scientific names of species into fixed $d$-dimensional embeddings.

\paragraph{\textbf{Training details:}}
We used $d'=128$ as the embedding dimension for each level of HIER-Embed, which when concatenated across the four levels produces the combined hierarchical embedding of $d=512$ dimensions. Phylo-Diffusion uses this $d$-dimensional embedding to condition LDMs through cross-attention in denoising the U-Net backbone and train LDMs without classifier-free guidance. We used VQGAN \cite{esser2021taming} as the backbone encoder-decoder to achieve the latent representations desired for LDMs with a downsampling factor of 4. All the models with different encoders are trained for 400k iterations, employing the best model checkpoint if convergence occurs early. Additional hyperparameters, such as learning rate, batch size, and U-Net architecture, are detailed in \Cref{app:hyperparameter}.

\begin{table}[t]
\centering
\caption{Quantitive comparison of generated images sampled using DDIM \cite{song2020denoising} (100 samples per class).}
\begin{tabular}{llcccc}
\hline
\textbf{Model Type} & \textbf{Method} & \textbf{FID ↓} & \textbf{IS ↑}  & \textbf{Prec. ↑} & \textbf{Recall ↑} \\ \hline
GAN & Phylo-NN & 28.08 & 2.35 & 0.625 & 0.084  \\
Diffusion & Class Conditional & 11.46 & 2.47 & 0.679 & 0.359   \\
Diffusion & Scientific Name & 11.76 & 2.43 & 0.683 & 0.332  \\
Diffusion & Phylo-Diffusion (ours) & 11.38 & 2.53 & 0.654 & 0.367 \\ \hline
\end{tabular}
\label{tab:fid}
\end{table}

\section{Results}

\subsection{Quality of Generated Images}
\Cref{tab:fid} compares the quality of generated images of baselines using the metrics of Fréchet Inception Distance (FID) score, Inception Score (IS), and Precision, Recall calculated in the feature space as proposed in \cite{kynkaanniemi2019improved}. Our results show that Phylo-Diffusion is at par with state-of-the-art generative models, achieving an FID of 11.38 compared to LDM's 11.46. We show a sample of generated images in \Cref{fig:samples}, with additional images provided in \Cref{app:additional-samples}. We also show the robustness of Phylo-Diffusion's results with varying numbers of phylogenetic levels and embedding dimensions in \Cref{app:ablations}.

\begin{table}[t]
\centering
\caption{Classification F1-Score on the 100 samples generated per class. The base classifier has an accuracy of \textit{85\%} on the test set.}
\begin{tabular}{lc}
\hline
\textbf{Method} &  \textbf{F1-Score (in \%) ↑} \\ \hline
Phylo-NN &  47.37 \\
Class Conditional &  81.99 \\
Scientific Name &  70.16 \\
Phylo-Diffusion (ours)  & 82.21 \\ \hline
\end{tabular}
\label{tab:classification}
\end{table}

\subsection{Classification Accuracy}
\label{sec:classifier}
We used a separate model for species classification, specifically a ResNet-18 model \cite{he2016deep} trained using the same training/validation split as Phylo-Diffusion. The primary objective behind building this classifier is to verify if images generated by Phylo-Diffusion contain sufficient discriminatory information to be classified as their correct species classes. \Cref{tab:classification}  compares the classification F1-scores over 100 samples generated by baseline conditioning schemes. We can see that the synthetic images generated by Phylo-Diffusion achieve the highest F1 score (82.21\%), which is quite close to the F1 score of the base classifier on the original test images (85\%). We present additional results showing the generalizability of Phylo-Diffusion in classifying generated images to unseen species in \Cref{app:ablations}.

\subsection{Matching Embedding Distances with Phylogenetic Distances}

\begin{figure*}[t]
\centering
\begin{subfigure}{0.24\textwidth}
\includegraphics[width=\textwidth]{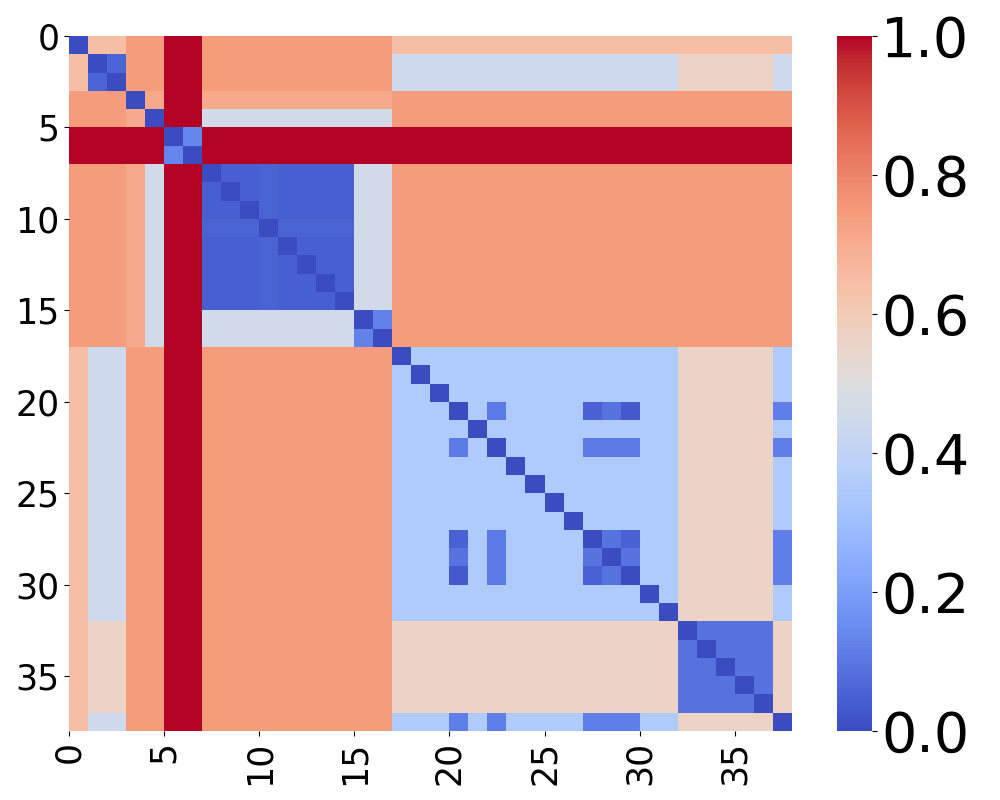}
\caption{GT Phylogenetic Distance}
\label{fig:dist_gt}
\end{subfigure}
\hfill
\begin{subfigure}{0.24\textwidth}
\includegraphics[width=\textwidth]{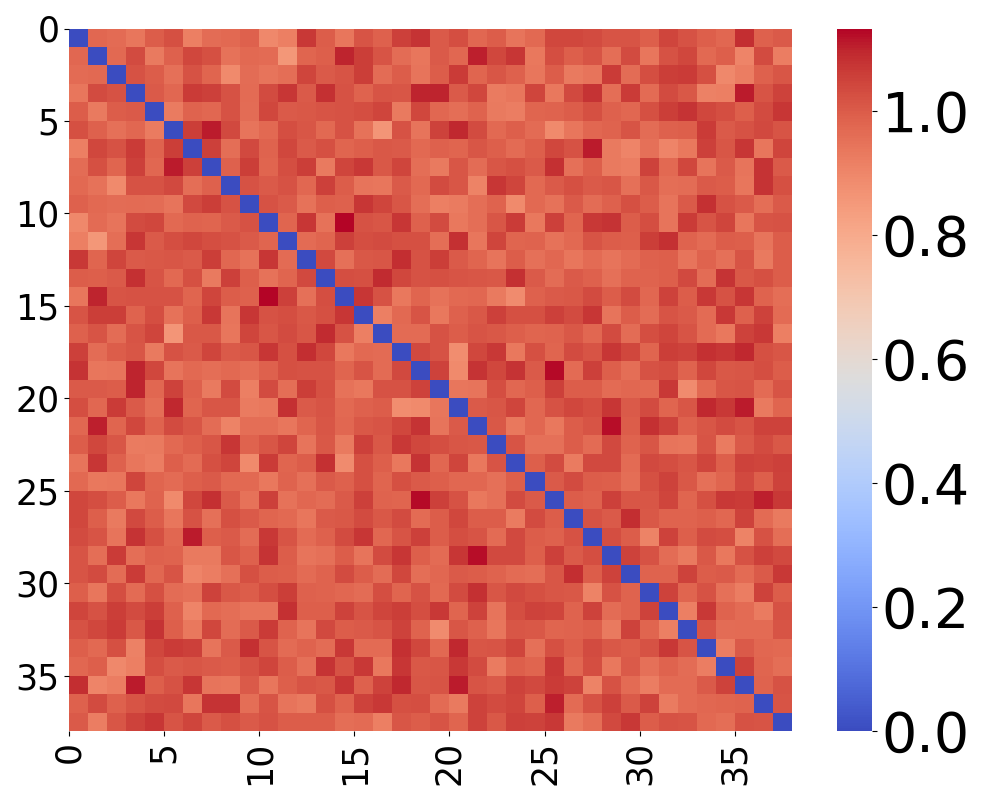}
\caption{Class Conditional Encoding }
\label{fig:dist_class_label}
\end{subfigure}
\hfill
\begin{subfigure}{0.24\textwidth}
\includegraphics[width=\textwidth]{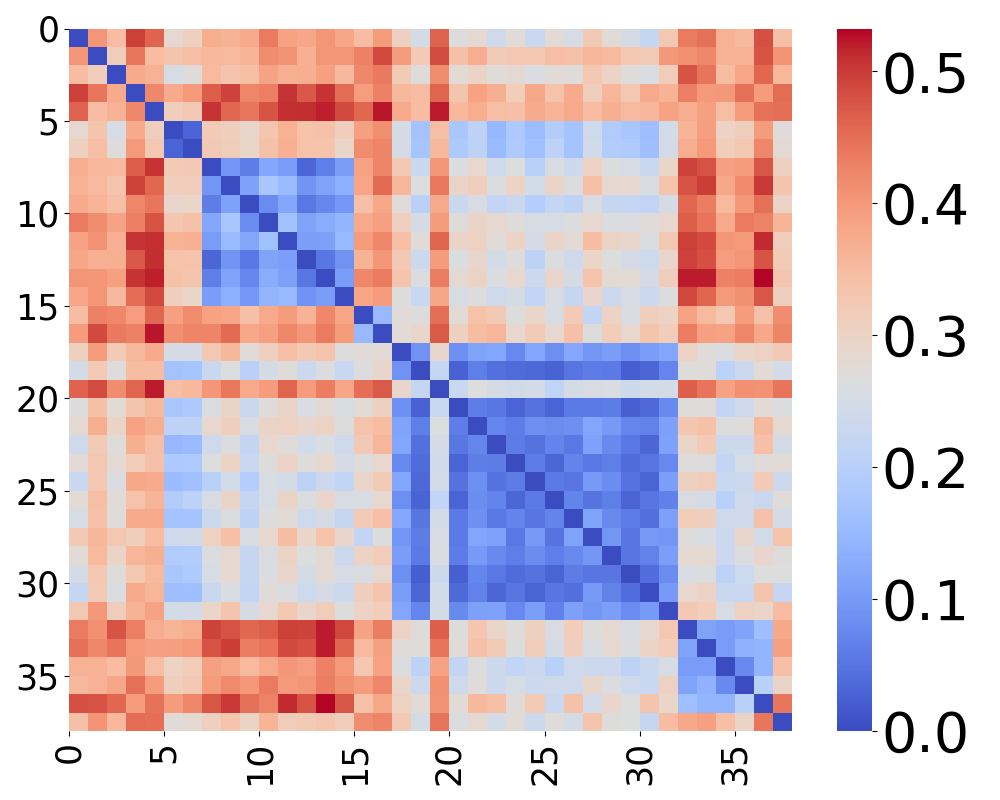}
\caption{Scientific Name Encoding}
\label{fig:dist_sci_clip}
\end{subfigure}
\hfill
\begin{subfigure}{0.24\textwidth}
\includegraphics[width=\linewidth]{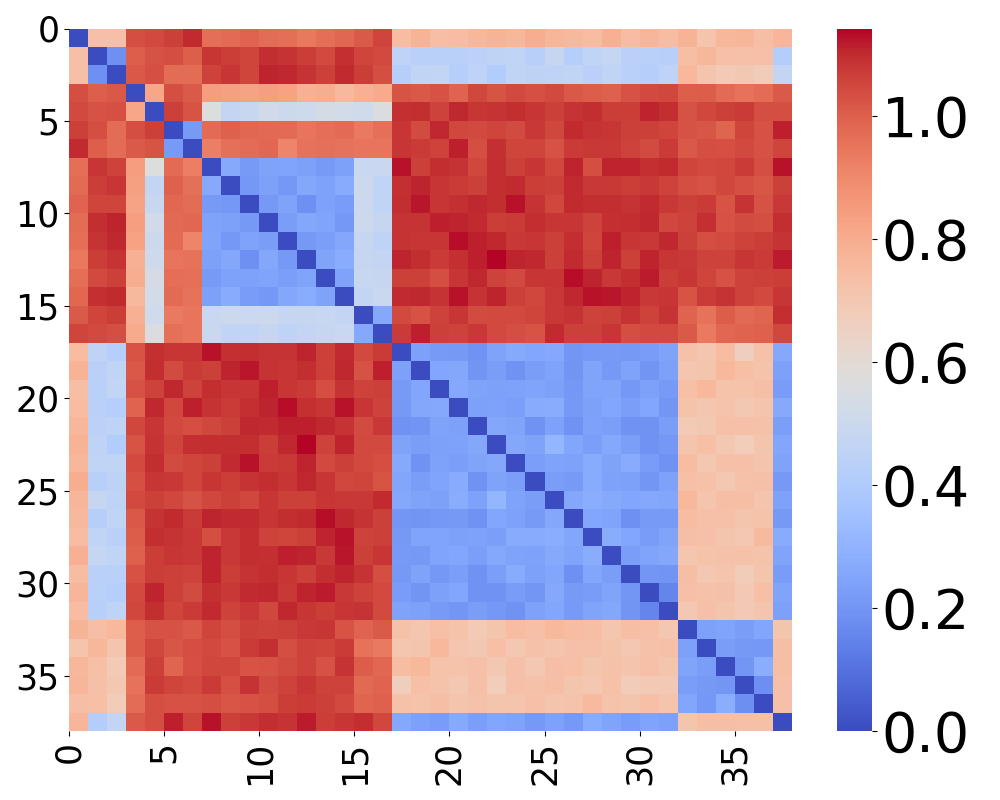}
\caption{Hierarchical Embedding}
\label{fig:dist_hier_lvl}
\end{subfigure}
\caption{Comparing Cosine distances in the embedding space of species for varying conditioning mechanisms.}
\label{fig:phylo_distance}
\end{figure*}

We investigate the quality of embeddings produced by baseline methods by comparing distances in the embedding space with the ground-truth (GT) phylogenetic distances computed from the tree of life, as illustrated in \Cref{fig:phylo_distance}.
Ideally, we expect distances in the embedding space of species pairs to be reflective of their phylogenetic distances. 
For Class Conditional, we can see that the distance matrix does not show any alignment with the GT phylogenetic distance matrix.
In the case of Scientific Name Encoding, the distance matrix exhibits notable similarities to the phylogenetic distances, thanks to the hierarchical nature of information contained in scientific names (i.e., \textit{genus}-name \& \textit{species}-name). However, one limitation of this encoding is its inability to capture inter-genus similarities or differences. In contrast, HIER-Embed shows a distance matrix that closely aligns with the GT phylogenetic distance matrix, validating its ability to preserve evolutionary distances among species in its embedding space.

\begin{figure}[t]
\centering
\includegraphics[width=1\linewidth]{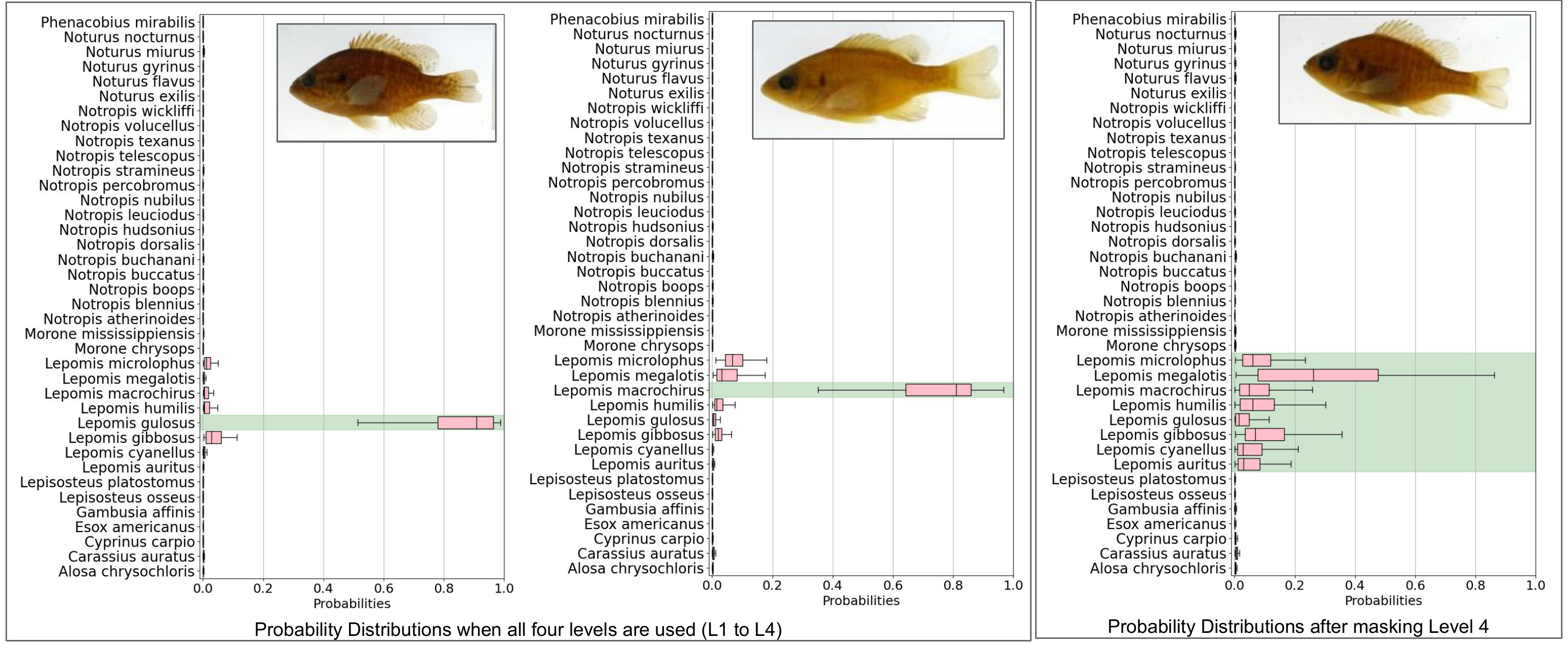}
\caption{Left: class probability distributions of images generated by using embeddings at all four levels for two species \textit{Lepomis gulosus} and \textit{Lepomis macrochirus} (shown in green) that are part of the same sub-tree till level 3. Right: class probability distributions of images generated by masking level 4 (descendant species that have common ancestry till level 3 are highlighted in green)
}
\label{fig:Logits}
\end{figure}

\subsection{Trait Masking Results}

To obtain classification probabilities or logits associated with generated images, we employ the classifier detailed in Section \ref{sec:classifier}. For the masked embeddings of subtrees at level 3, defined as $[\eone, \etwo, \ethree, \znoise ]$, we analyze the logits of generated images and compare them with those generated without masking. \Cref{fig:Logits} demonstrates that for a specific subtree, in this case \textit{Lepomis}, logits for species within the subtree are higher compared to those for species outside it. This outcome aligns with the expectation that Phylo-Diffusion, when provided with information up to Level 3, can capture overarching characteristics of all species within the given subtree. Additionally, \Cref{fig:Logits} presents probability distributions for species within the \textit{Lepomis} subtree when the full set of hierarchical encodings $[\eone, \etwo, \ethree, \efour]$ is provided. It demonstrates that the probabilities are significantly higher for the targeted class, as intended for image generation. After masking, we observe that the generated images are very similar and capture common features of the \textit{Lepomis} genus. For all our calculations and plots, we generate 100 images for each subtree and node.
\Cref{app:trait_masking} contains additional histograms that detail logit distributions across all different subtrees at each level, offering comprehensive insights into how the model discriminates and learns the hierarchical structure across different levels and nodes.

\begin{figure}[t]
\centering
\includegraphics[width=0.6\linewidth]{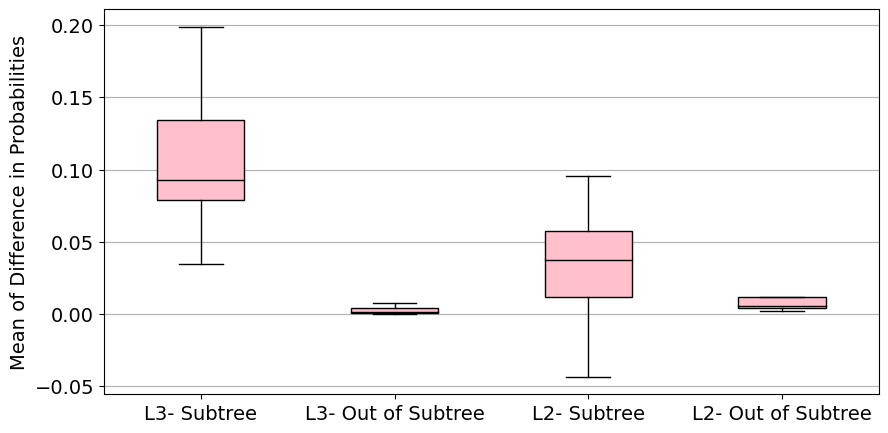}
\caption{Box plot for the mean of difference in probabilities for species within the subtree and out of subtree for level-3 and level-2.}
\label{fig:distribution}
\end{figure}

\paragraph{\textbf{Quantitative Evaluation of Probability Distrubtions:}}
\label{sec:prob_change}
To quantitatively evaluate the ability of Phylo-Diffusion to capture hierarchical information and show desired changes in probability distributions after masking, we compute the following metrics. Let us denote the set of all species in the data as $\mathcal{S}$ and for a given sub-tree at an internal node $i$ of level $l$, let us denote the subset of descendant species as $\mathcal{S}_{i}^l = \{S_1, S_2, \ldots, S_n\}$. We first compute the reference probabilities $P_{ref}$ of every species before masking (\ie, by using all four level embeddings). Let us denote the probability of predicting a generated image using all four embeddings of a descendant species $S_j \in \mathcal{S}_{i}^l$ into species class $S_k$ as $P_{S_j}(S_k)$. The reference probability of a species $S_k$ can then be given as: 
\begin{equation}
 P_{ref}(S_k) = \begin{cases}
    \frac{1}{|\mathcal{S}_i^l|-1}\sum_{S_j \in \mathcal{S}_i^l \setminus S_k} P_{S_j}(S_k),& \text{if } S_k \in \mathcal{S}_i^l,\\
    \frac{1}{|\mathcal{S}_i^l|}\sum_{S_j \in \mathcal{S}_i^l} P_{S_j}(S_k),             & \text{if } S_k \not \in \mathcal{S}_i^l.
\end{cases}  
\end{equation}
Note that when $S_k$ is part of the sub-tree, \ie, $S_k \in \mathcal{S}_i^l$,  $P_{ref}(S_k)$ is computed by averaging over $|\mathcal{S}_i^l|-1$ probability values since we exclude the case when $S_k$ is used to generate the images. On the other hand, when $S_k$ is outside of the sub-tree, \ie,  $S_k \not \in \mathcal{S}_i^l$, we average over all  $|\mathcal{S}_i^l|$ probability values. Given these reference probabilities values before masking, we can compute the change in probability of predicting species $S_k$ after masking as $P_{diff}(S_k) = P_{mask}(S_k) - P_{ref}(S_k)$, where $P_{mask}(S_k)$ is the probability of predicting a generated image after masking to $S_k$. We expect $P_{diff}$ to be larger for descendant species $S_{k} \in \mathcal{S}_i^l$ compared to species that are outside of the sub-tree because of the dispersion of probabilities in a sub-tree as a consequence of masking.  We thus compute the average $P_{diff}$ for species that belong to subtree $\mathcal{S}_i^l$ as  $P_{diff}^{sub}(i,l)$ and species that are outside the subtree $\mathcal{S}_i^l$ as $P_{diff}^{out}(i,l)$.

\Cref{fig:distribution} shows the box plot of $P_{diff}^{sub}$ and $P_{diff}^{out}$ across internal nodes at levels 2 and 3. We observe that species within the subtree exhibit a more pronounced increase in probabilities compared to species outside the subtree, aligning with our expectations. Notably, this trend is consistently observed across both levels 2 and 3. More details of class-wise probability distribution shifts for each level are provided in \Cref{app:trait_masking}.
The outcomes of these experiments affirm that Phylo-Diffusion effectively identifies unique features at Levels 2 and 3, and captures shared features or traits of any chosen subtree at the internal nodes of the phylogeny.

\subsection{Trait Swapping Results}

\begin{figure}[t]
\centering

\begin{subfigure}[b]{0.7\textwidth}
   \includegraphics[width=1\linewidth]{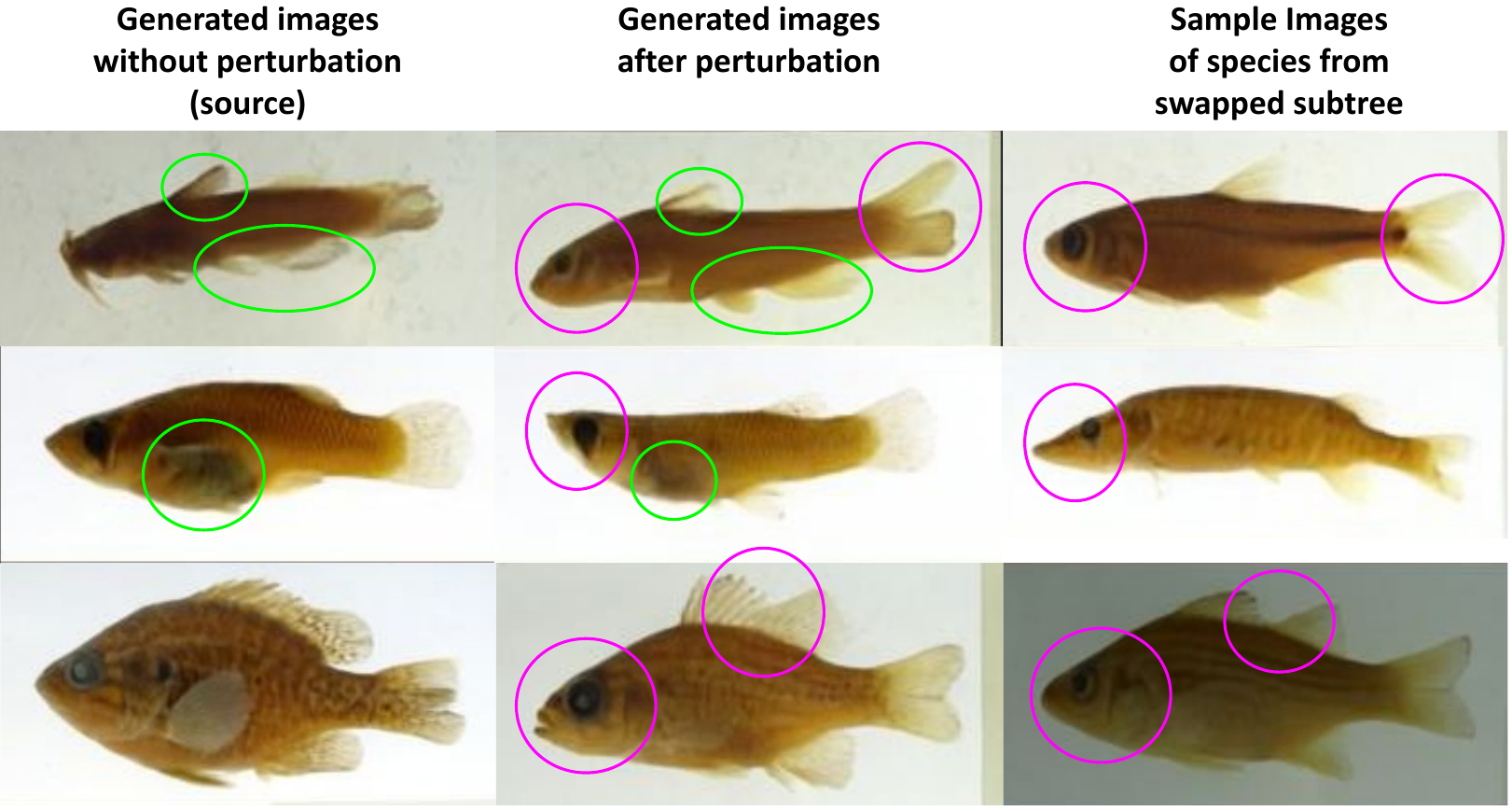}
   \caption{Examples of traits swapping for species at level-2 (first two rows) and level-3 (last row). The order of species from top to bottom is \textit{Noturus exilis} swapped with \textit{Notropis} and \textit{Gambusia affinis} swapped with \textit{Esox americanus}. The third row shows trait swapping at level-3 for \textit{Lepomis gulosus} swapped with \textit{Morone}.}
   \label{fig:swapping1}
\end{subfigure}

\begin{subfigure}[b]{0.7\textwidth}
   \includegraphics[width=1\linewidth]{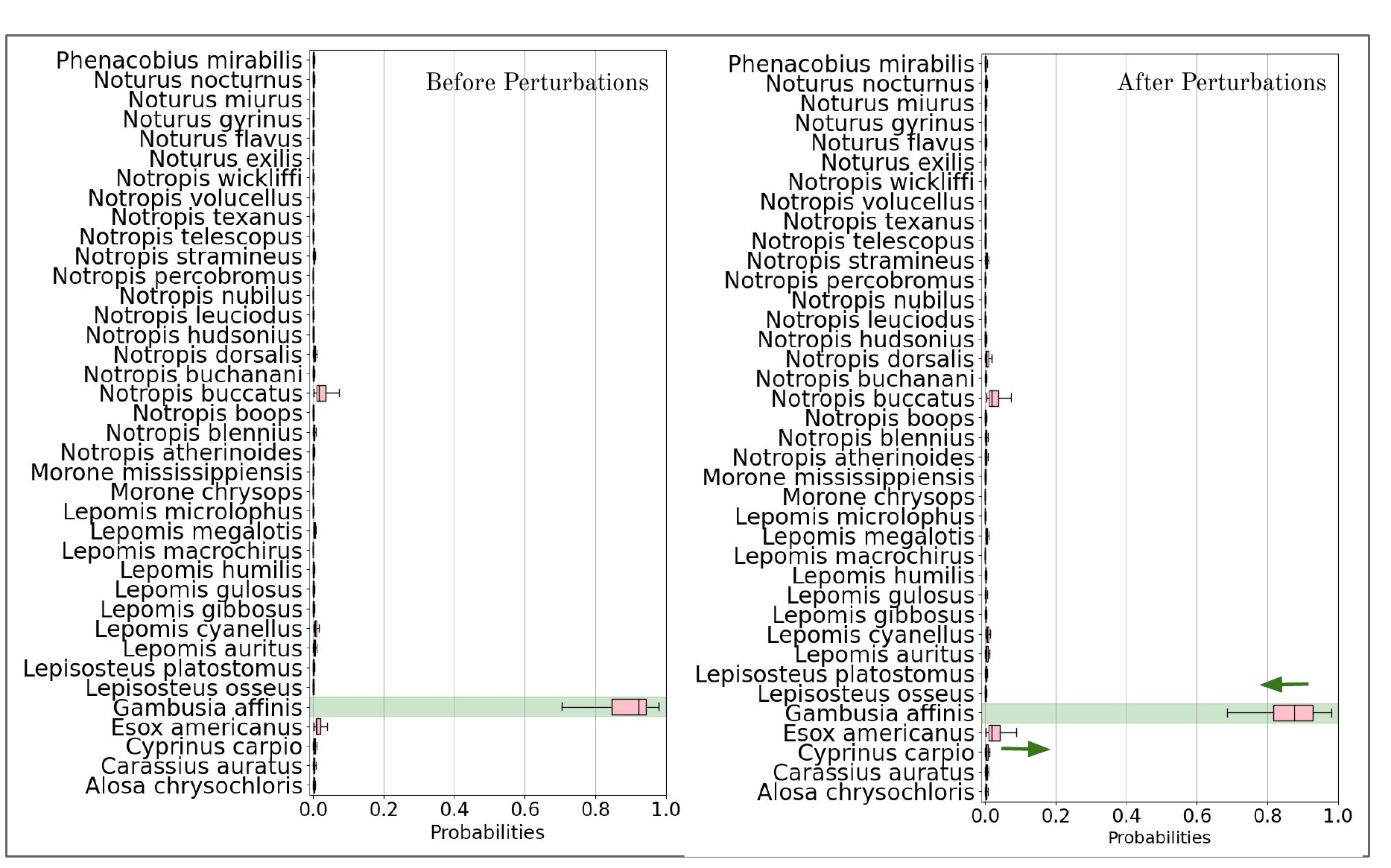}
   \caption{For Row 2 of Figure \ref{fig:swapping1}, we show that the probability distribution of \textit{Gambusia affinis} decreases after the swapping traits at level-2, with an increase in the probability distribution of \textit{Esox americanus}.}
   \label{fig:swapping2}
\end{subfigure}

\caption{Examples of trait swapping results.}
\label{fig:substitution}
\end{figure}

Figure \ref{fig:substitution} shows examples of trait swapping results enabled by Phylo-Diffusion. For the first example (first row of \Cref{fig:swapping1}), we swap the level-2 embedding of source species \textit{Noturus exilis} with level-2 embedding of its sibling group \textit{Notropis/ Carassius}. The goal here is to discover traits of \textit{Noturus exilis} inherited at level-2 that differentiate it from other groups of species that branched out at this point of time in evolution. We can see that the generated images of the perturbed embedding (center) exhibit the absence of barbels (whiskers) highlighted in purple, while the caudal (or tail) fin is beginning to fork (or split), a trait adopted from \textit{Notropis} (right). In contrast, other fins such as the dorsal, pelvic, and anal fins highlighted in green remain similar to those of the source species, \textit{Noturus exilis} (left). This suggests that at level-2, \textit{Notropis} and \textit{Noturus} species diverged by developing differences in two distinct traits, barbels and forked caudal fins while keeping other traits intact.

For the second example (\Cref{fig:swapping1}, row 2), we swap level-2 information of \textit{Gambusia affinis} (left) with that of \textit{Esox americanus} (right). The generated images of the perturbed embedding (center) exhibit a more pointed head highlighted in purple, and a slimmer body shape resembling \textit{Esox americanus}. Notably, the perturbed species retains discoloration at the bottom from the source species highlighted in green. \Cref{fig:swapping2} presents probability distributions (or logits) of \textit{Gambusia affinis} before and after trait swapping using the classifier detailed in Section \ref{sec:classifier}. We observe a slight decrease in logits for \textit{Gambusia affinis} and an increase in logits for \textit{Esox americanus}, consistent with our expectations. In the third row of \Cref{fig:swapping1}, we swap level-3 information of \textit{Lepomis gulosus} (left) with that of \textit{Morone} genus (right). The resulting images from the perturbed embedding (center) capture the horizontal line pattern characteristic of \textit{Morone} genus, and the dorsal fin highlighted in purple begins to split. All these examples suggest novel scientific hypotheses about differences in evolutionary traits acquired by species at different ancestry levels, which can be validated by biologists in subsequent studies.
Note that our experiments are most effective at levels near the species nodes, specifically at levels 2 \& 3, since phylogenetic signal is known to diminish as we move toward the root of the tree \cite{harmon2010early, pennell2015model}.
Additional visualizations of trait swapping results are provided in \Cref{app:trait_swapping}.

\begin{figure}[t]

\begin{subfigure}{0.47\textwidth}
    \includegraphics[width=\textwidth]{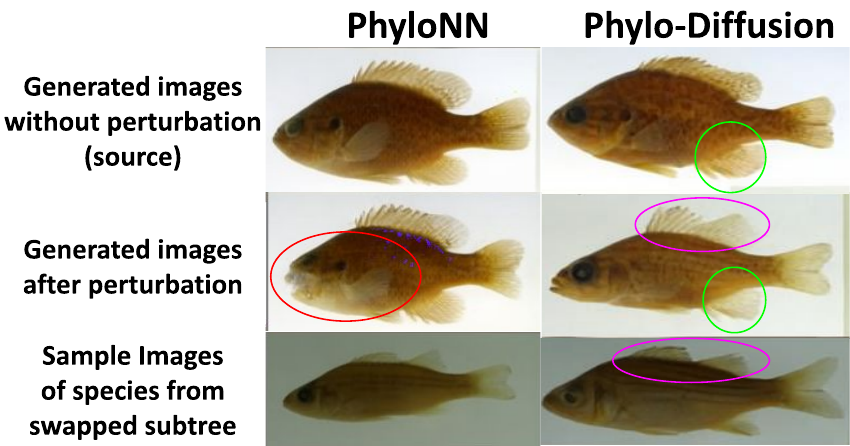}
    \caption{Swapping Level 3 traits for \textit{Lepomis gulosus} with \textit{Morone}}
    \label{fig:Phylo1}
\end{subfigure}
\begin{subfigure}{0.52\textwidth}
    \includegraphics[width=\textwidth]{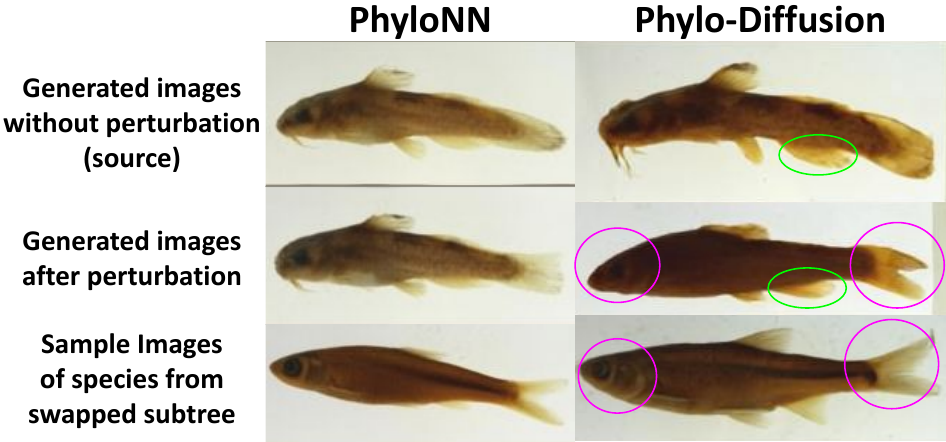}
    \caption{Swapping Level 2 traits for \textit{Notorus mirurus} with \textit{Notropis}}
    \label{fig:Phylo2}
\end{subfigure}
\caption{Comparing Phylo-NN with Phylo-Diffusion for examples of trait swapping.}
\label{fig:phylo_nn_comapre}
\end{figure}

\paragraph{\textbf{Comparisions with Phylo-NN:}}
\Cref{fig:phylo_nn_comapre} compares trait swapping results of Phylo-Diffusion and Phylo-NN for the same set of example species. \Cref{fig:Phylo1} shows trait swapping at level-3 for the source species of \textit{Lepomis gulosus} (top) and target sub-tree of the \textit{Morone} genus (bottom). In Phylo-NN, images generated by perturbing the Imageome sequences appear blurry (red circle), while Phylo-Diffusion effectively captures the splitting of dorsal fin (purple circle) and the horizontal stripe pattern of the \textit{Morone} genus, while maintaining the fin structure of \textit{Lepomis gulosus} (green circle). Similarly, \Cref{fig:Phylo2} compares trait swapping for \textit{Noturus miurus} (top) with the target sub-tree of \textit{Notropis} genus (bottom) at level-2. For Phylo-NN, the perturbed images are almost identical to the source species. However, Phylo-Diffusion shows visible trait differences such as the absence of barbels and the caudal (or tail) fin beginning to fork or split (purple circle), which are traits picked from the target sub-tree of \textit{Notropis} genus. Note that we had considered the same target sub-tree in \Cref{fig:swapping1} row 1 and observed similar trait differences in the generated images after perturbation, further validating the ability of Phylo-Diffusion to discover consistent evolutionary traits. We provide additional results comparing Phylo-Diffusion and Phylo-NN trait swapping results in \Cref{app:phylonn-compare}.
\section{Conclusions and Future Work}
In this work, we introduced Phylo-Diffusion, a novel framework for discovering evolutionary traits from images by structuring the embedding space of diffusion models using tree-based knowledge. 
In the future, our approach can be extended to work on other applications involving image data linked with phylogenies or pedigrees.
Our work also has limitations that need to be addressed in future research. For example, while our current work is limited to discretized trees with a fixed number of levels, future works can focus on discovering evolutionary traits at every internal node of the phylogenetic tree with varying levels without performing any discretization. Future works can also attempt to capture convergent changes in evolution, i.e., changes that occur repeatedly in different branches of the tree, and perform ancestral state reconstruction with uncertainty estimates.

\section*{Acknowledgments}
This research is supported by National Science Foundation (NSF) awards for the HDR
Imageomics Institute (OAC-2118240). We are thankful for the support of computational resources provided by the Advanced Research Computing (ARC) Center at Virginia Tech.

This manuscript has been authored by UT-Battelle, LLC, under contract DE-AC05-00OR22725 with the US Department of Energy (DOE). The US government retains and the publisher, by accepting the article for publication, acknowledges that the US government retains a nonexclusive, paid-up, irrevocable, worldwide license to publish or reproduce the published form of this manuscript, or allow others to do so, for US government purposes. DOE will provide public access to these results of federally sponsored research in accordance with the DOE Public Access Plan (
https://www.energy.gov/doe-public-access-plan).


%
%
\bibliographystyle{splncs04}
\bibliography{main}

\clearpage
\appendix
\clearpage
\section*{Supplementary Materials}
Here is a summary of additional details and experiments included in the Appendices.

\begin{enumerate}
    \item{\cref{app:hyperparameter}: Hyperparameter Settings and Training Details}
    \item{\cref{app:phylogy}: Details about Phylogenetic Tree}
    \item{\cref{app:trait_masking}: Additional Details about Trait Masking Experiments}
    \item{\cref{app:trait_swapping}: Additional Examples for Trait Swapping Experiments}
    \item{\cref{app:phylonn-compare}: Additional Comparisons with PhyloNN}
    \item{\cref{app:additional-samples}: Additional Samples of Generated Images}
    \item{\cref{app:ablations}: Ablation Results}
    \item{\cref{app:cub_results}: CUB Dataset Results}

\end{enumerate}

\section{Hyperparameter Settings and Training Details} \label{app:hyperparameter}

\Cref{tab:hyperparameter} lists all the hyperparameters for the models trained. We used cross-attention as the conditioning mechanism for all the models and all the models were trained from scratch. At the inference stage, we used DDIM \cite{song2020denoising}sampling with 200 steps. For computing metrics like FID, IS, \etc, we use ADM's \cite{dhariwal2021diffusion} TensorFlow evaluation script.

\begin{table}
\centering
\caption{Hyperparameter settings  of the baselines and Phylo-Diffusion.}
\label{tab:hyperparameter}
\begin{tabular}{lccc}
\toprule
       \textbf{Model}    & Class Conditional &    Scientific Name & Phylo-Diffusion \\   
\midrule
    \textit{z}-shape    &  64 × 64 × 3 &     64 × 64 × 3 &  64 × 64 × 3 \\
    Diffusion Steps    & 1000  & 1000  & 1000\\
    Noise Schedule    & linear  & linear  & linear\\
    Model Size    & 469M  &  902M & 469M \\
    Channels    & 224  & 224  & 224\\
    Depth    &  2 &  2 & 2 \\
    Channel Multiplier    &   1,2,3,4 &   1,2,3,4 &  1,2,3,4 \\
    Attention resolutions    & 32, 16, 8  &  32, 16, 8 & 32, 16, 8 \\
    Number of Heads    & 32  & 32  & 32 \\
    Dropout    & -  & -  & -\\
    Batch Size    & 8  & 8 & 8\\
    Iterations    & 400k  &  400k & 400k\\
    Learning Rate    & 4e-5  &  4e-5 & 4e-5 \\
    Scale    & 1  & 1  & 1\\
    Embedding Dimension    &  1 x 512 &  77 x 768 & 1 x 512\\
    Transformers Depth    &  1 &  1 & 1 \\
\bottomrule
\end{tabular}
\end{table}

All diffusion models require about 7 days to train on a single A100 GPU for both bird and fish datasets. Inference throughput is 0.9 samples/sec using DDIM with 200 steps computed over generating 100 images per class. We do not have any additional overheads in training and inference time compared to LDMs.

\section{Details about Phylogenetic Tree} \label{app:phylogy}
\begin{figure}
\centering
\includegraphics[width=1\linewidth]{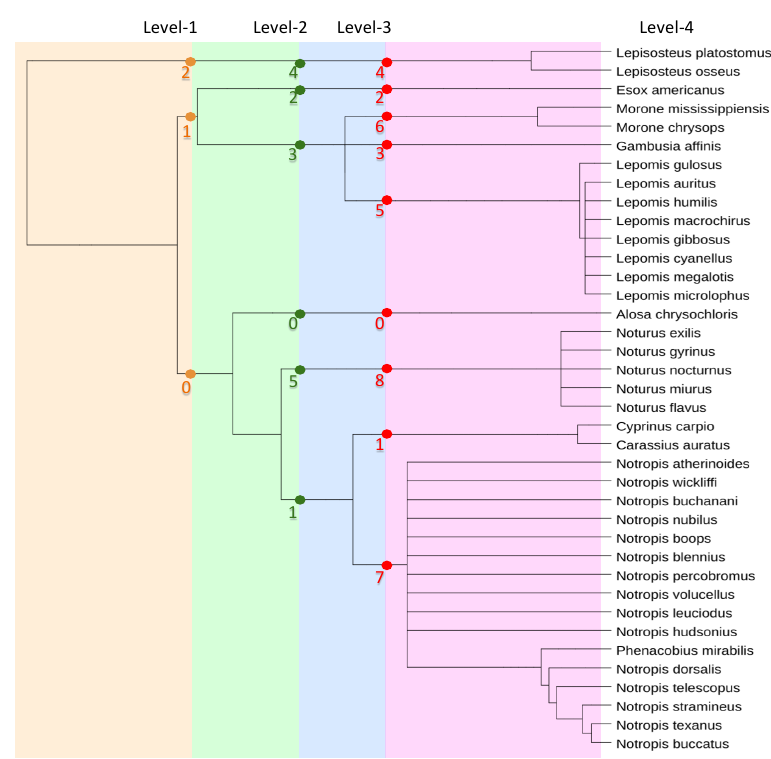}
\caption{Phylogeny tree for fishes for all 38 species. Filled circles show nodes of the subtrees defined at each of the four levels after discretization.}
\label{fig:phylogeny}
\end{figure}

\Cref{fig:phylogeny} shows the phylogeny tree for all the species in the fish dataset along with the information of the four discrete levels used in our study (marked by different colored circles). \Cref{tab:groupings} and \ref{tab:groupings2} list out all the groupings (subtrees) made after discretizing the tree into four levels where the fourth level is the species itself.

\begin{table}[hbt!]
\caption{Phylogenetic groupings of fish species included in this study at different ancestry levels.}
\label{tab:groupings}
\begin{tabular}{p{0.1\linewidth}|p{0.17\linewidth}|p{0.73\linewidth}}
\toprule
       Level  & Node at level   &    Species groupings \\
\midrule
   3                    &   Node 0         & \textit{Alosa chrysochloris} \\[0.2cm]
                       &   Node 1                 & \textit{Carassius auratus}, \textit{Cyprinus carpi} \\[0.2cm]
                       &   Node 2       & \textit{Esox americanus} \\[0.2cm]
                       &   Node 3      & \textit{Gambusia affinis} \\[0.2cm]
                       &   Node 4                  & \textit{Lepisosteus osseus}, \textit{Lepisosteus platostomus} \\[0.2cm]
                       &   Node 5                  & \textit{Lepomis auritus}, \textit{Lepomis cyanellus}, \textit{Lepomis gibbosus}, \textit{Lepomis gulosus}, \textit{Lepomis humilis}, \textit{Lepomis macrochirus}, \textit{Lepomis megalotis}, \textit{Lepomis microlophus} \\[0.2cm]
                       &   Node 6                  & \textit{Morone chrysops}, \textit{Morone mississippiensis} \\[0.2cm]
                       &   Node 7                  & \textit{Notropis atherinoides}, \textit{Notropis blennius}, \textit{Notropis boops}, \textit{Notropis buccatus}, \textit{Notropis buchanani}, \textit{Notropis dorsalis}, \textit{Notropis hudsonius}, \textit{Notropis leuciodus}, \textit{Notropis nubilus}, \textit{Notropis percobromus}, \textit{Notropis stramineus}, \textit{Notropis telescopus}, \textit{Notropis texanus}, \textit{Notropis volucellus}, \textit{Notropis wickliffi}, \textit{Phenacobius mirabilis} \\[0.2cm]
                       &   Node 8                 & \textit{Noturus exilis}, \textit{Noturus flavus}, \textit{Noturus gyrinus}, \textit{Noturus miurus}, \textit{Noturus nocturnus} \\

  \midrule
  2           &   Node 0         & \textit{Alosa chrysochloris} \\[0.2cm]
                         &   Node 1               &  \textit{Carassius auratus}, \textit{Cyprinus carpio}, \textit{Notropis atherinoides}, \textit{Notropis blennius}, \textit{Notropis boops}, \textit{Notropis buccatus}, \textit{Notropis buchanani}, \textit{Notropis dorsalis}, \textit{Notropis hudsonius}, \textit{Notropis leuciodus}, \textit{Notropis nubilus}, \textit{Notropis percobromus}, \textit{Notropis stramineus}, \textit{Notropis telescopus}, \textit{Notropis texanus}, \textit{Notropis volucellus}, \textit{Notropis wickliffi}, \textit{Phenacobius mirabilis}\\[0.2cm]
                        &   Node 2                & \textit{Esox americanus} \\[0.2cm]
                        &   Node 3                & \textit{Gambusia affinis}, \textit{Lepomis auritus}, \textit{Lepomis cyanellus}, \textit{Lepomis gibbosus}, \textit{Lepomis gulosus}, \textit{Lepomis humilis}, \textit{Lepomis macrochirus}, \textit{Lepomis megalotis}, \textit{Lepomis microlophus}, \textit{Morone chrysops}, \textit{Morone mississippiensis} \\[0.2cm]
                        &   Node 4                & \textit{Lepisosteus osseus}, \textit{Lepisosteus platostomus} \\[0.2cm]
                        &   Node 5                & \textit{Noturus exilis}, \textit{Noturus flavus}, \textit{Noturus gyrinus}, \textit{Noturus miurus}, \textit{Noturus nocturnus} \\

\bottomrule
\end{tabular}
\end{table}

\begin{table}[ht]
\caption{Phylogenetic groupings of species included in this study at different ancestry levels (continued from \cref{tab:groupings}) }
\label{tab:groupings2}
\begin{tabular}{p{0.1\linewidth}|p{0.17\linewidth}|p{0.73\linewidth}}
\toprule
       Level    & Node at level &    Species groupings \\
\midrule
     1         &   Node 0        &  \textit{Alosa chrysochloris}, \textit{Carassius auratus}, \textit{Cyprinus carpio}, \textit{Notropis atherinoides}, \textit{Notropis blennius}, \textit{Notropis boops}, \textit{Notropis buccatus}, \textit{Notropis buchanani}, \textit{Notropis dorsalis}, \textit{Notropis hudsonius}, \textit{Notropis leuciodus}, \textit{Notropis nubilus}, \textit{Notropis percobromus}, \textit{Notropis stramineus}, \textit{Notropis telescopus}, \textit{Notropis texanus}, \textit{Notropis volucellus}, \textit{Notropis wickliffi}, \textit{Noturus exilis}, \textit{Noturus flavus}, \textit{Noturus gyrinus}, \textit{Noturus miurus}, \textit{Noturus nocturnus}, \textit{Phenacobius mirabilis}  \\[0.2cm]
     &   Node 1 & \textit{Esox americanus}, \textit{Gambusia affinis}, \textit{Lepomis auritus}, \textit{Lepomis cyanellus}, \textit{Lepomis gibbosus}, \textit{Lepomis gulosus}, \textit{Lepomis humilis}, \textit{Lepomis macrochirus}, \textit{Lepomis megalotis}, \textit{Lepomis microlophus}, \textit{Morone chrysops}, \textit{Morone mississippiensis}\\[0.2cm]
     &   Node 2 & \textit{Lepisosteus osseus}, \textit{Lepisosteus platostomus} \\[0.2cm]
\bottomrule
\end{tabular}
\end{table}

\section{Additional Details about Trait Masking Experiments}\label{app:trait_masking}

\subsubsection{Additional Visualizations of Changes in Probability Distributions after Masking:} 
\begin{figure}[ht]
\centering
\includegraphics[width=1\linewidth]{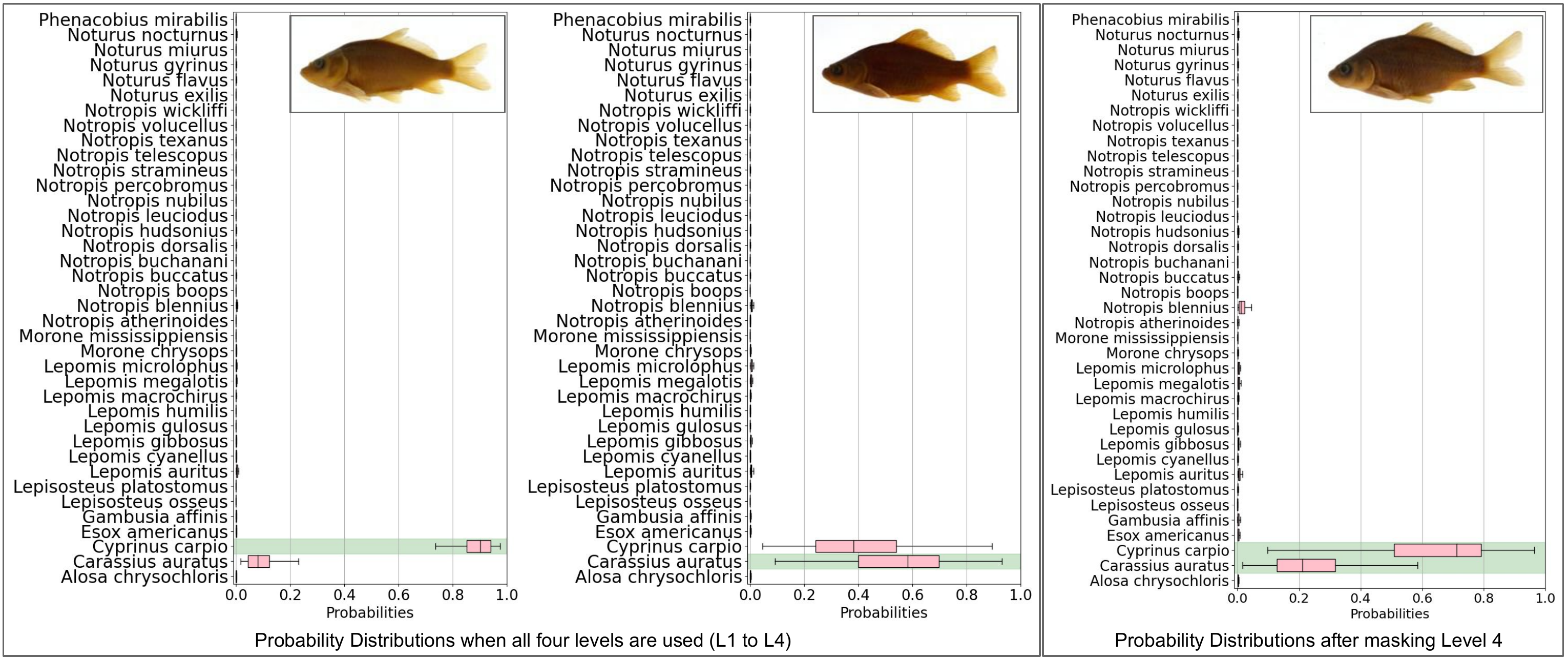}
\caption{Left: class probability distributions of images generated by using embeddings at all four levels for two species \textit{Cyprinus carpio} and \textit{Carassius auratus} (shown in green) that are part of the same sub-tree till level 3. Right: class probability distributions of images generated by masking level 4 (descendant species that have common ancestry till level 3 are highlighted in green)}
\label{fig:level3-class1}
\end{figure}

\begin{figure}[ht]
\centering
\includegraphics[width=1\linewidth]{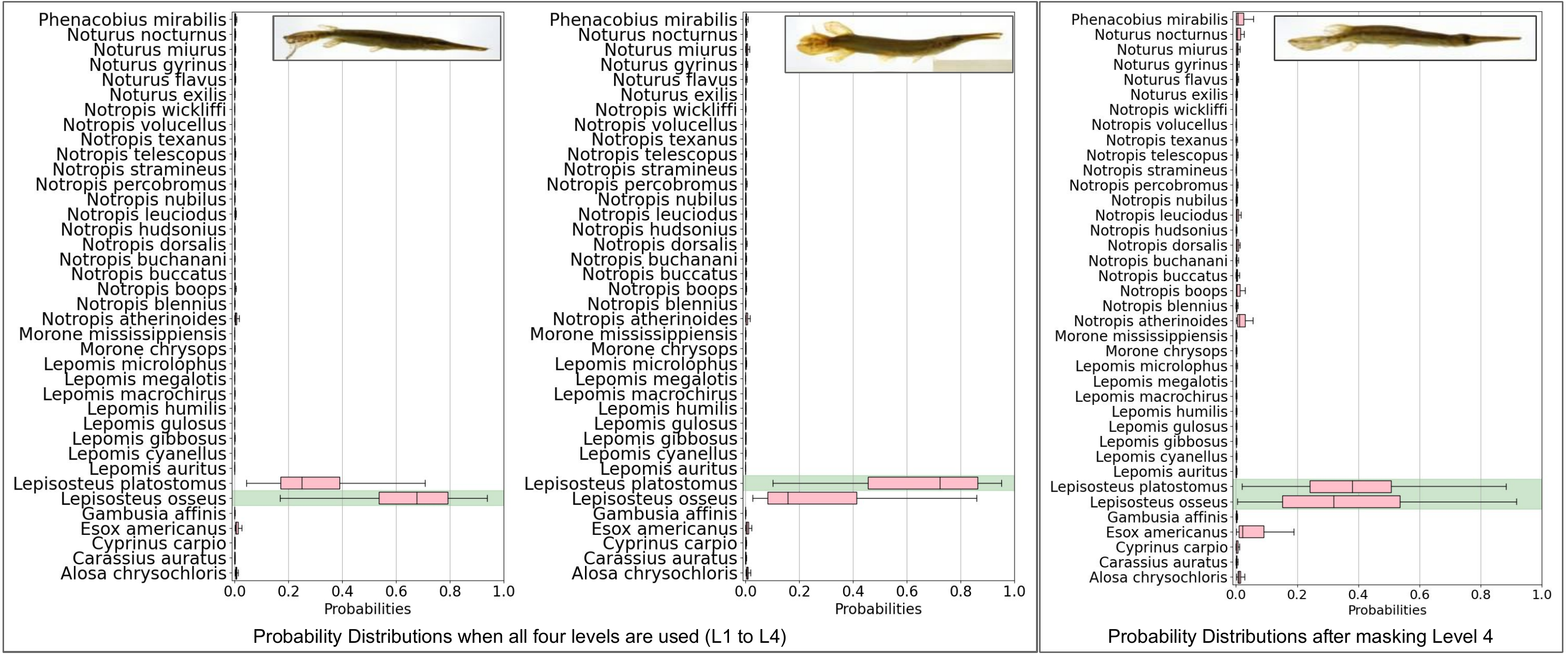}
\caption{Left: class probability distributions of images generated by using embeddings at all four levels for two species \textit{Lepisosteus osseus} and \textit{Lepisosteus platostomus} (shown in green) that are part of the same sub-tree till level 3. Right: class probability distributions of images generated by masking level 4 (descendant species that have common ancestry till level 3 are highlighted in green)}
\label{fig:level3-class4}
\end{figure}

\begin{figure}[ht]
\centering
\includegraphics[width=1\linewidth]{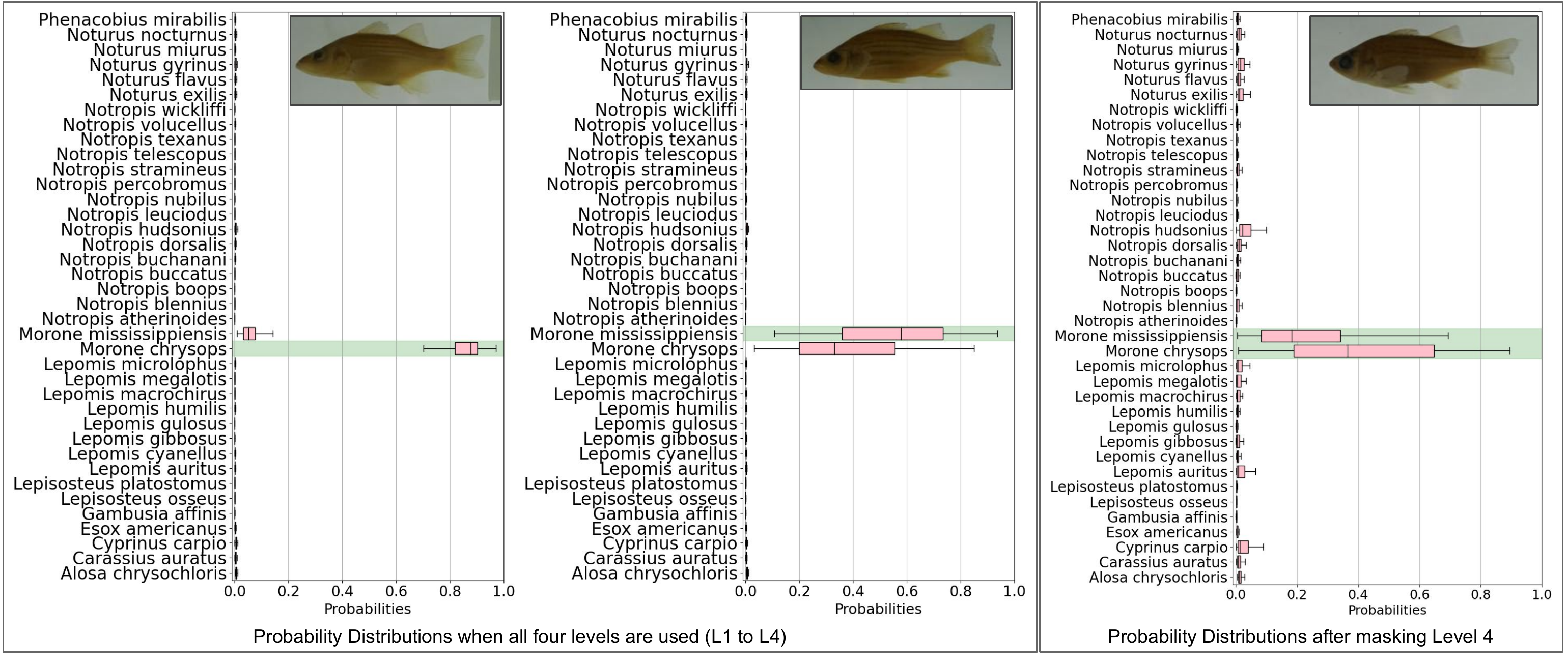}
\caption{Left: class probability distributions of images generated by using embeddings at all four levels for two species \textit{Morone chrysops} and \textit{Morone mississippiensis} (shown in green) that are part of the same sub-tree till level 3. Right: class probability distributions of images generated by masking level 4 (descendant species that have common ancestry till level 3 are highlighted in green)}
\label{fig:level3-class6}
\end{figure}

\begin{figure}[t]
\centering
\includegraphics[width=1\linewidth]{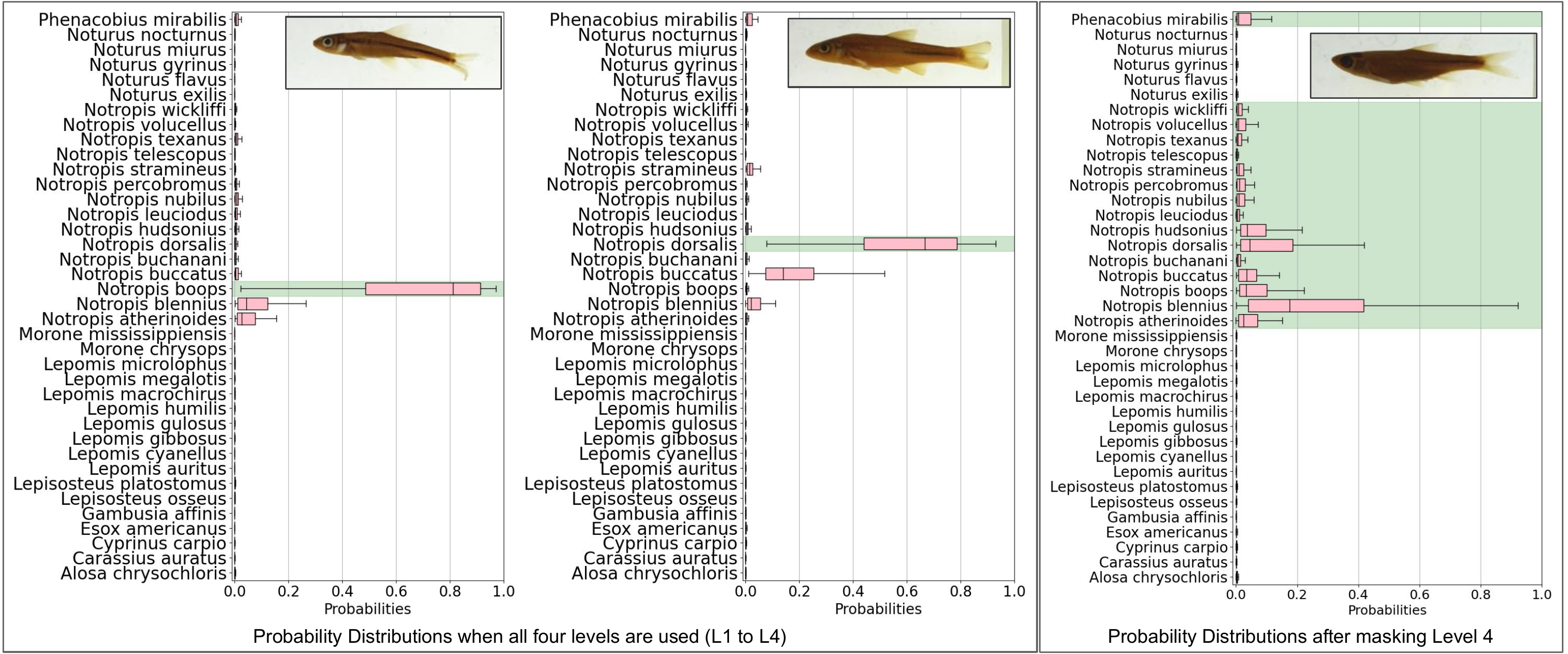}
\caption{Left: class probability distributions of images generated by using embeddings at all four levels for two species \textit{Notropis boops} and \textit{Notropis dorsalis} (shown in green) that are part of the same sub-tree till level 3. Right: class probability distributions of images generated by masking level 4 (descendant species that have common ancestry till level 3 are highlighted in green)}
\label{fig:level3-class7}
\end{figure}

\begin{figure}
\centering
\includegraphics[width=1\linewidth]{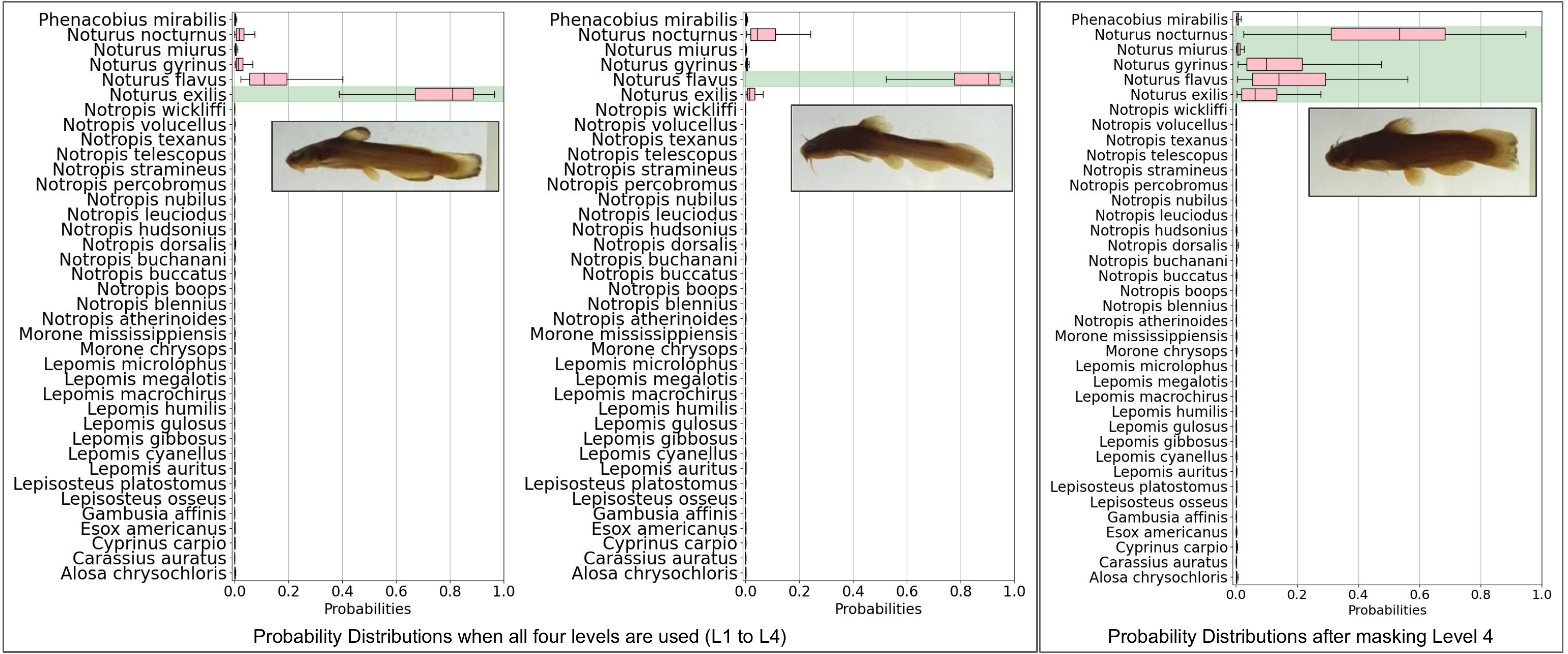}
\caption{Left: class probability distributions of images generated by using embeddings at all four levels for two species \textit{Noturus exilis} and \textit{Noturus falvus} (shown in green) that are part of the same sub-tree till level 3. Right: class probability distributions of images generated by masking level 4 (descendant species that have common ancestry till level 3 are highlighted in green)}
\label{fig:level3-class8}
\end{figure}

\begin{figure}
\centering
\includegraphics[width=1\linewidth]{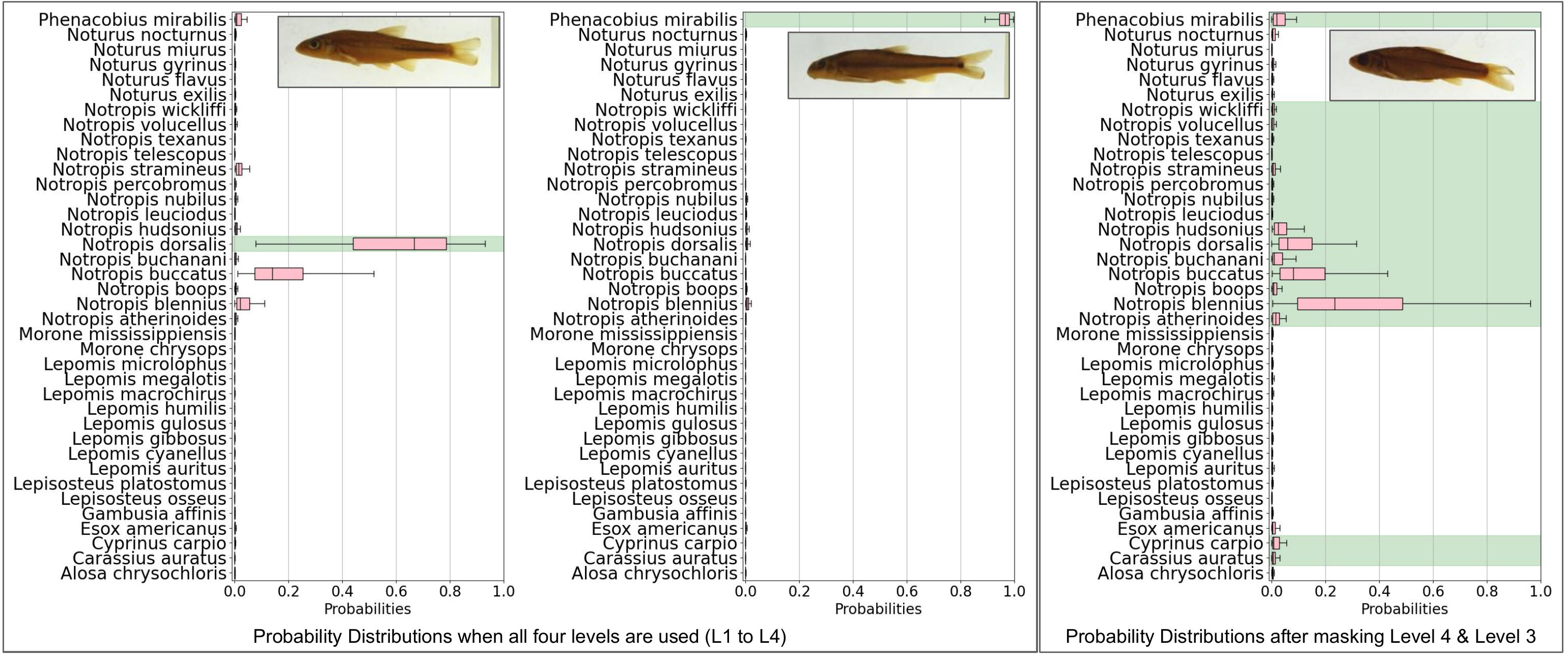}
\caption{Left: class probability distributions of images generated by using embeddings at all four levels for two species \textit{Notropis dorsalis} and \textit{Phenacobius mirabilis} (shown in green) that are part of the same sub-tree till level 2. Right: class probability distributions of images generated by masking level 3 and level 4 (descendant species that have common ancestry till level 2 are highlighted in green)}
\label{fig:level2-class1}
\end{figure}

\begin{figure}
\centering
\includegraphics[width=1\linewidth]{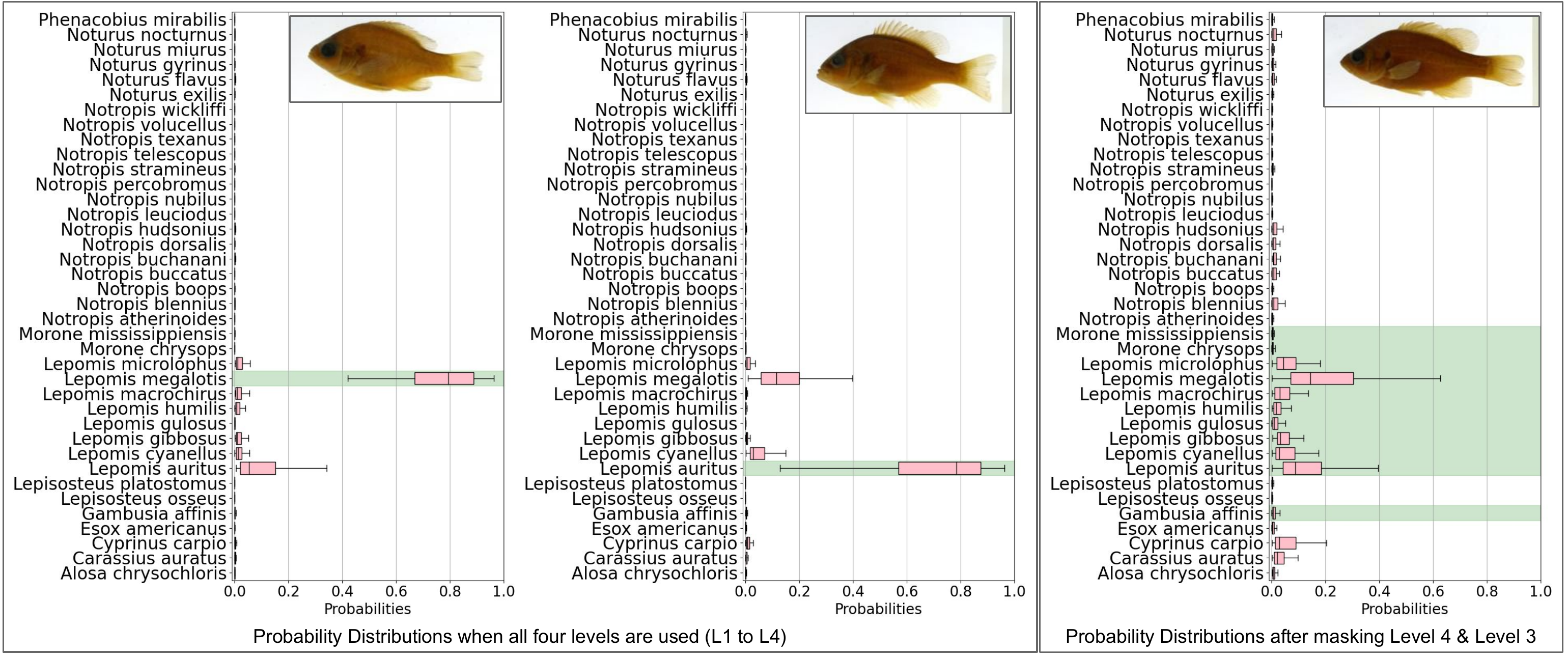}
\caption{Left: class probability distributions of images generated by using embeddings at all four levels for two species \textit{Lepomis megalotis} and \textit{Lepomis auritus} (shown in green) that are part of the same sub-tree till level 2. Right: class probability distributions of images generated by masking level 3 and level 4 (descendant species that have common ancestry till level 2 are highlighted in green)}
\label{fig:level2-class3}
\end{figure}

\begin{figure}
\centering
\includegraphics[width=1\linewidth]{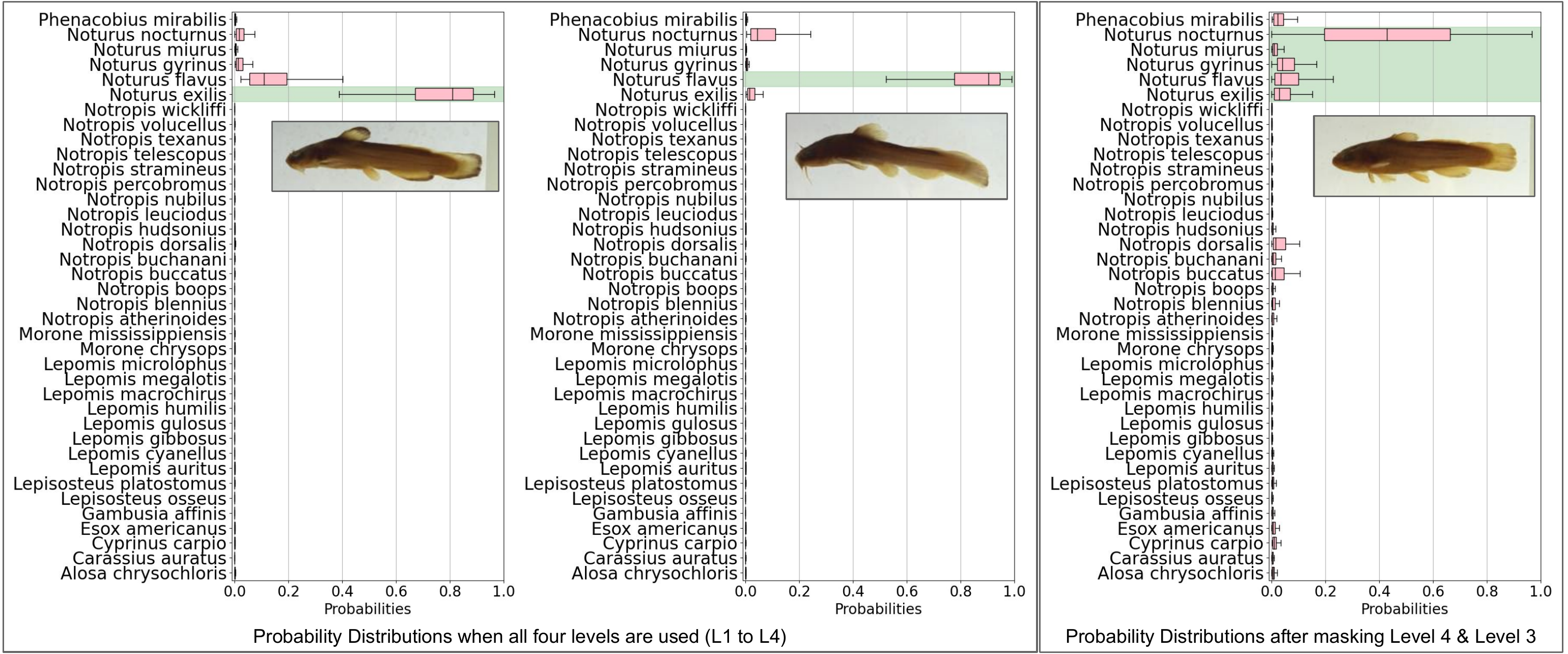}
\caption{Left: class probability distributions of images generated by using embeddings at all four levels for two species \textit{Noturus exilis} and \textit{Noturus flavus} (shown in green) that are part of the same sub-tree till level 2. Right: class probability distributions of images generated by masking level 3 and level 4 (descendant species that have common ancestry till level 2 are highlighted in green)}
\label{fig:level2-class5}
\end{figure}

\Cref{fig:level3-class1}, \ref{fig:level3-class4}, \ref{fig:level3-class6}, \ref{fig:level3-class7} and \ref{fig:level3-class8} show additional examples of changes in probability distributions when level-4 information is replaced with \textit{noise}. In each figure, the first two plots display probability distributions (or logits) of images generated using embeddings from all four levels, \ie $[\eone, \etwo, \ethree, \efour]$, of two representative species sharing a common ancestry up to level-3 (highlighted in green). We show that the logits are higher for the targeted species as expected. The third plot logits after masking level-4 embeddings, leading to a dispersion of probabilities across all descendant species within the subtree up to level-3 (highlighted in green). The only exception is \Cref{fig:level3-class7}, where there is some skewness in the logits of descendant species, which is likely due to the data imbalance across classes at higher levels of the tree and also due to biases in the classifier (classifier test accuracy is 85\% as reported in \Cref{sec:classifier}). In this case, the classifier sometimes misclassifies \textit{Notropis boops} as \textit{Notropis blennius} in the first plot and \textit{Notropis dorsails} as \textit{Notropis buccatus} in the second plot. Consequently, the third plot for the \textit{Notropis} subtree shows a higher probability for \textit{Notropis blennius}. Similarly, \Cref{fig:level2-class1}, \ref{fig:level2-class3} and \ref{fig:level2-class5} provide examples of trait masking where both Level 3 and 4 are replaced with \textit{noise}, \ie $[\eone, \etwo, \znoise, \znoise]$. We observe a similar trend in the dispersion of probabilities across all descendant species within the same subtree at level-2. In all trait masking visualizations, we consistently observe that logits of generated images of species within the subtree (highlighted in green) are higher than for species outside the subtree. This demonstrates Phylo-Diffusion's ability to effectively capture hierarchical information at various levels of the phylogenetic tree.

\clearpage

\subsubsection{Additional Quantitative Results of Trait Masking:} Tables \ref{tab:level3_logits} and \ref{tab:level2_logits} show the change in probabilities for different nodes at levels 3 and 2, respectively. We can see that indeed $P_{diff}^{sub}$ is larger than $P_{diff}^{out}$ for all internal nodes at levels 2 and 3 (except node 5 at level 2), indicating that Phylo-Diffusion is capturing the necessary hierarchical information required for the dispersion of probabilities after masking. \Cref{fig:distribution} in the main paper shows the box plots of Tables \ref{tab:level3_logits} and \ref{tab:level2_logits}. \Cref{tab:level_logits} summarizes this information by showing the average $P_{diff}^{sub}$ and $P_{diff}^{out}$ for all nodes at a given level. It is important to note that for this experiment, we focus on nodes that have more than one species in the defined subtree.

\begin{table}
\centering
\caption{Average change in probability distributions for every node at Level 3.}
\begin{tabular}{ccc}
\hline
\textbf{Node} &  \textbf{    Subtree} & \textbf{    Out-of-Subtree} \\ \hline
Node 1 & 0.1988 & 0.0018 \\
Node 4 & 0.0952 & 0.0051 \\
Node 5 & 0.0753 & 0.0007 \\
Node 6 & 0.0903 & 0.0076 \\
Node 7 & 0.0346 & 0.0006 \\
Node 8 & 0.1472 & 0.0003 \\ \hline
\end{tabular}
\label{tab:level3_logits}
\end{table}

\begin{table}
\centering
\caption{Average change in probability distributions for every node at Level 2.}
\begin{tabular}{ccc}
\hline
\textbf{Node} &  \textbf{    Subtree} & \textbf{    Out-of-Subtree} \\ \hline
Node 1 & 0.0299 & 0.0023 \\
Node 3 & 0.0449 & 0.0062 \\
Node 4 & 0.0952 & 0.0051 \\
Node 5 & -0.0434 & 0.0292 \\ \hline
\end{tabular}
\label{tab:level2_logits}
\end{table}

\begin{table}
\centering
\caption{Average change in probability distributions across all nodes at a certain level.}
\begin{tabular}{ccc}
\hline
\textbf{Levels} &  \textbf{    Subtree} & \textbf{    Out-of-Subtree} \\ \hline
Level 3 & 0.1070 & 0.0027 \\
Level 2 & 0.0316 & 0.0107 \\ \hline
\end{tabular}
\label{tab:level_logits}
\end{table}

\clearpage

\section{Additional Examples for Trait Swapping Experiments}\label{app:trait_swapping}

\subsubsection{Additional Visualizations of Trait Swapping Experiments:} 
\begin{figure}
\centering
\includegraphics[width=1\linewidth]{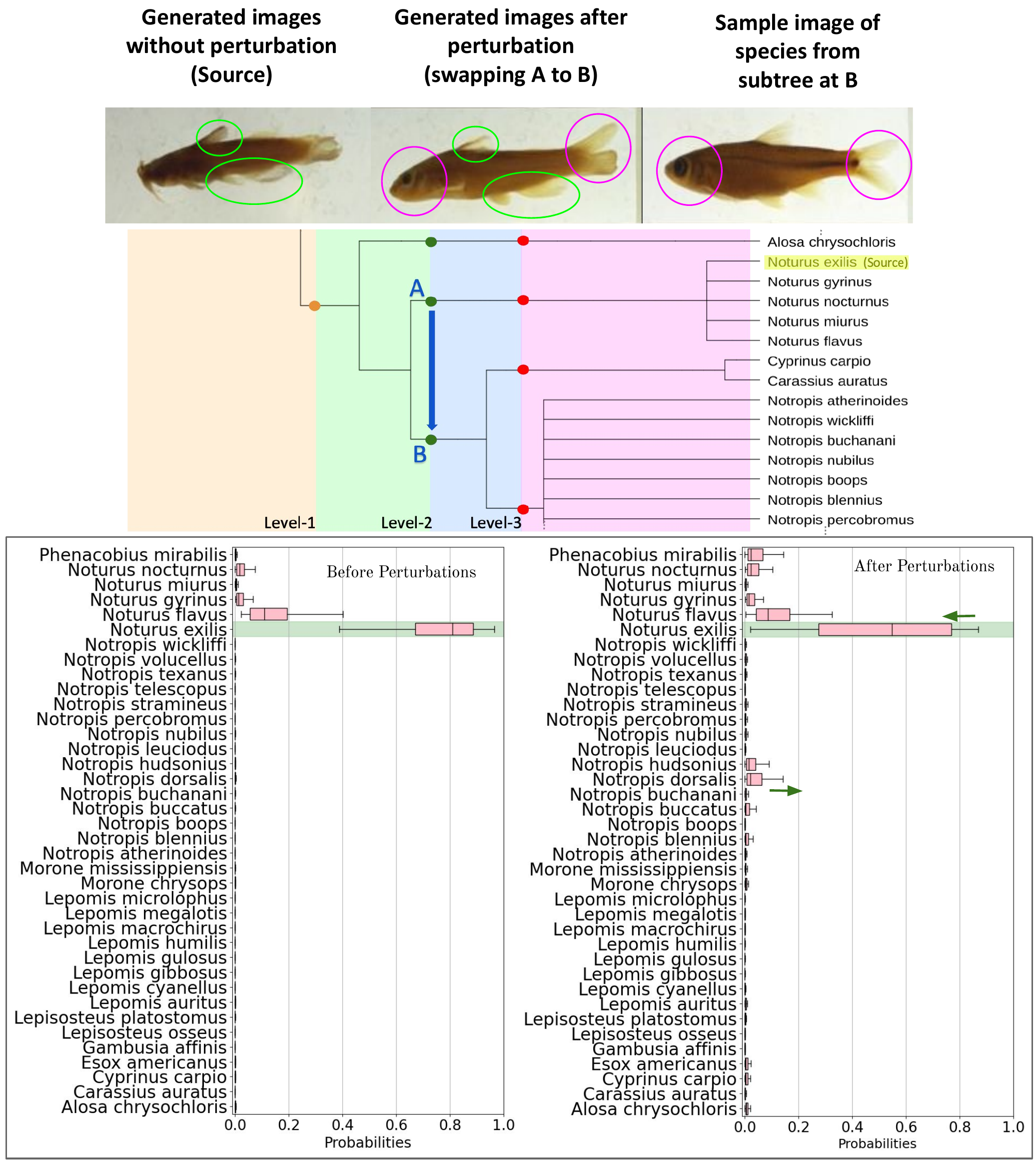}
\caption{Visualization of changes in traits after swapping information at Level 2 (Node \textit{A}) for \textit{Noturus exilis} (left) with its sibling subtree at Node \textit{B}(right) to generate perturbed species (center). Traits shared with the source species are outlined in green, whereas those shared with the sibling subtree at Node B are outlined in pink.}
\label{fig:trait_swapping1}
\end{figure}

\Cref{fig:trait_swapping1} illustrates trait swapping for the source species \textit{Noturus exilis} (left), where the information at Level-2 is swapped with that of a sibling subtree at Node \textit{B} (right). The image in the center is generated using the trait swapped embedding. This visualization of the perturbed species helps us study the trait changes that would have branched out at level-2 between Node \textit{A} and Node \textit{B}.In the generated image (center),  we observe the absence of barbels(whiskers), and the caudal fin (tail) is getting forked (or split) highlighted in pink, which are traits adopted from species in the subtree at \textit{B} (\textit{Notropis}). Whereas other fins like the dorsal, pelvic, and anal fin still resemble the source species \textit{Noturus exilis} highlighted in green. The same is also reflected in the change of probability distribution after perturbations; the probability distribution of source species \textit{Noturus exilis} decreases and the probability of it being a \textit{Notropis} increases slightly.

\begin{figure}
\centering
\includegraphics[width=1\linewidth]{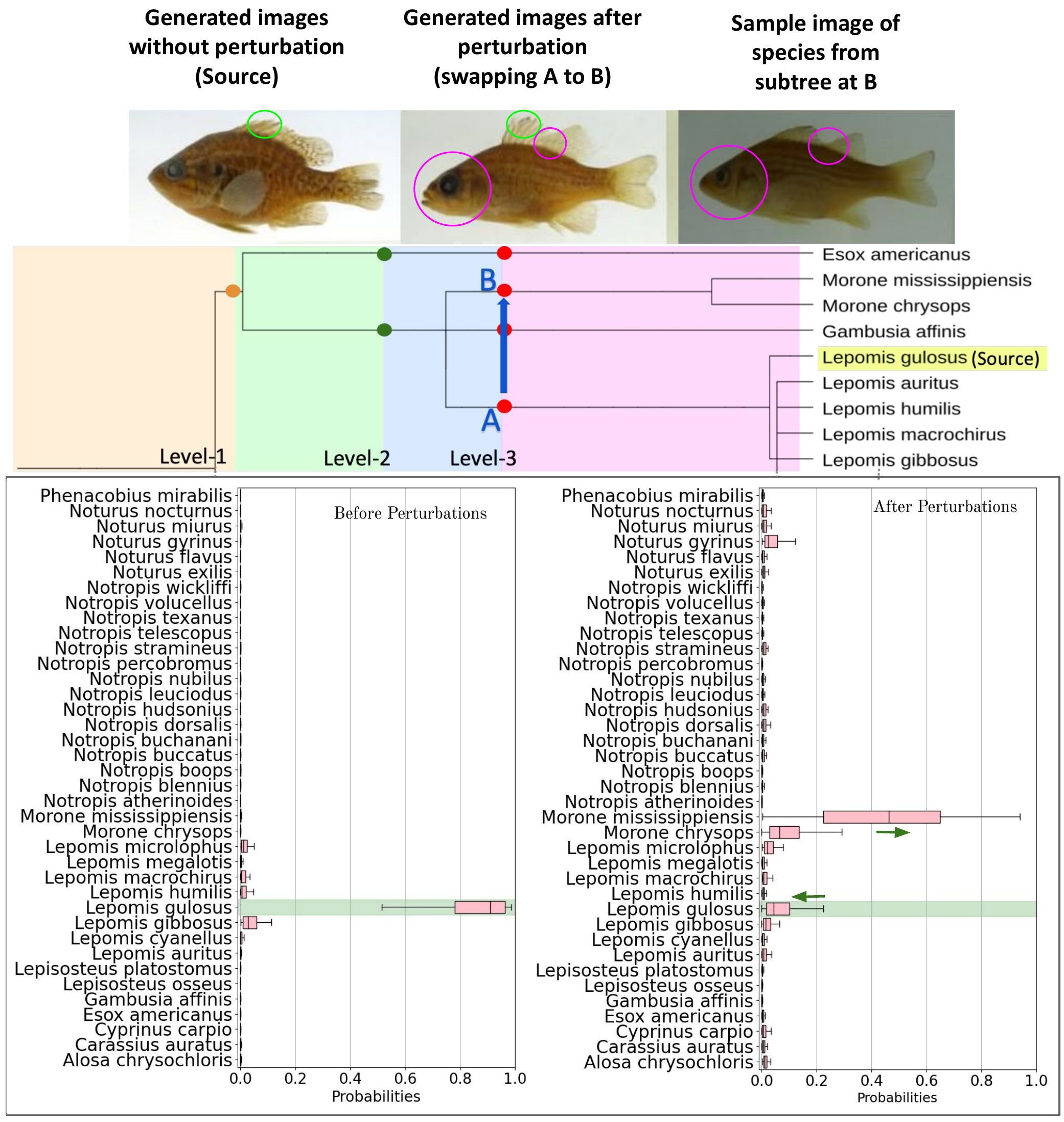}
\caption{Visualization of changes in traits after swapping information at Level 2 (Node \textit{A}) for \textit{Lepomis gulosus} (left) with its sibling subtree at Node \textit{B}(right) to generate perturbed species (center). Traits shared with the source species are outlined in green, whereas those shared with the sibling subtree at Node B are outlined in pink.}
\label{fig:trait_swapping2}
\end{figure}

Similarly in \Cref{fig:trait_swapping2}, for the studied species \textit{Lepomis gulosus} (left), the information at Level 3 is swapped with subtree at Node B (\textit{Morone}). The perturbed species generated (center) captures traits from both lineages. The spotted pattern in the body and fins is retained from \textit{Lepomis} (Node \textit{A}) but these spots now start to follow the horizontal stripes pattern observed in \textit{Morone} (species at Node \textit{B}). Additionally, the dorsal fin highlighted in pink starts to split into two, with the left half retaining the spiny structure from \textit{Lepomis} highlighted in green. This observation suggests that the species at Node \textit{A} and \textit{B} possess distinct traits at this level-3 branching node. The same is reflected in the shift in probability distributions towards species in Node \textit{B} after trait swapping.

\clearpage
\section{Additional Comparisons with PhyloNN}\label{app:phylonn-compare}

\begin{figure}
\centering
\includegraphics[width=1\linewidth]{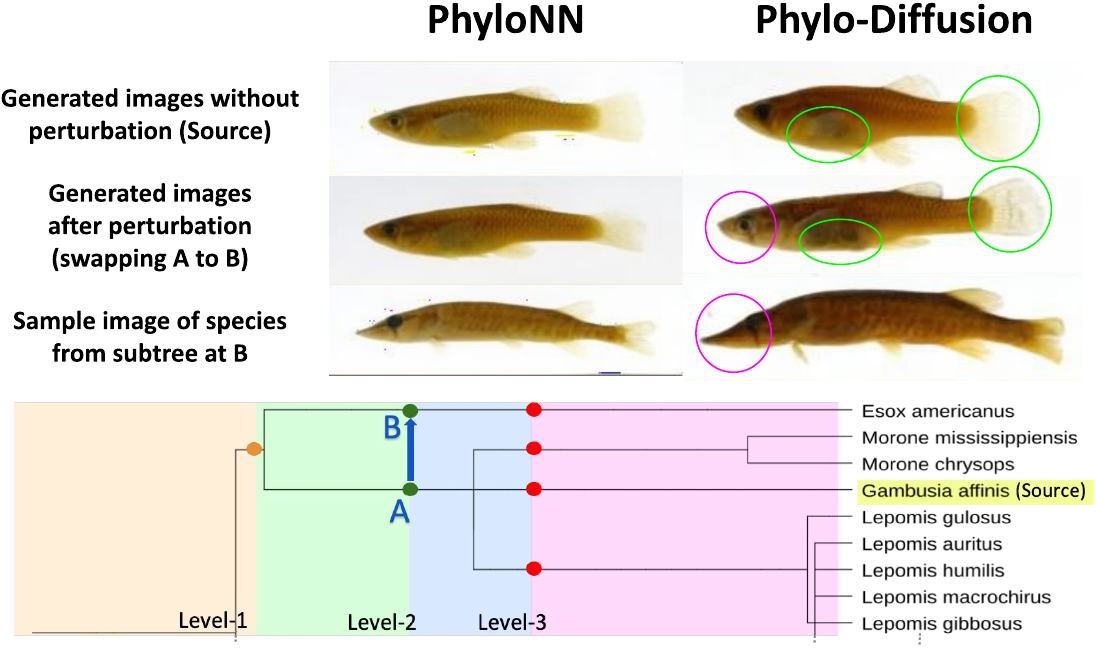}
\caption{Comparison of PhyloNN with Phylo-Diffusion (ours) for trait swapping where the Level-2 information (Node \textit{A}) of \textit{Gambusia affinis} is swapped with its sibling subtree at Node \textit{B} to generate perturbed species (center). Traits shared with the source species are outlined in green, whereas those shared with the sibling subtree at Node B are outlined in pink.}
\label{fig:gambusia_l2}
\end{figure}

\Cref{fig:gambusia_l2} compares the trait swapping experiment for Phylo-Diffusion with the PhyloNN baseline, where level-2 information of \textit{Gambusia affinis} is replaced with that of \textit{Esox americanus}. In the highlighted pink circle, the face of the image generated after perturbations (center) becomes more pointed, and the body shape flattens to resemble \textit{Esox americanus}. This perturbed image also retains traits like the caudal (tail) fin and the black-spotted pattern towards the bottom (highlighted in green) from the source species, \textit{Gambusia affinis}. The differences observed with Phylo-Diffusion are notable, whereas the PhyloNN generates a perturbed image nearly identical to the original, showing no significant changes.

\begin{figure}[t]
\centering
\includegraphics[width=1\linewidth]{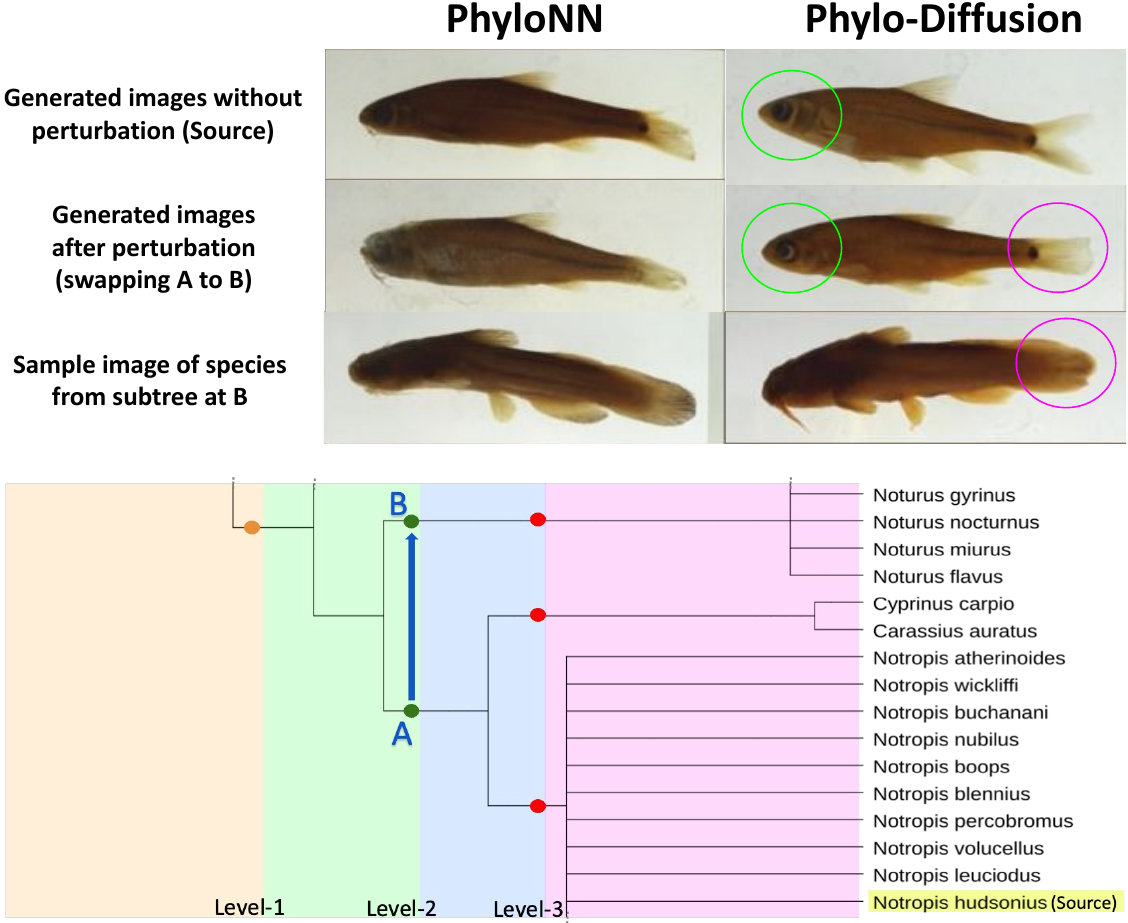}
\caption{Comparison of PhyloNN with Phylo-Diffusion (ours) for trait swapping where the Level-2 information (Node \textit{A}) of \textit{Notropis husonius} is swapped with its sibling subtree at Node \textit{B} to generate perturbed species (center). Traits shared with the source species are outlined in green, whereas those shared with the sibling subtree at Node B are outlined in pink.}
\label{fig:notropis_hudsonius_l2}
\end{figure}

Similarly, \Cref{fig:notropis_hudsonius_l2} shows a comparison after replacing level-2 information of \textit{Notropis husonius} with that of \textit{Noturus}. For Phylo-Diffusion, the caudal (tail) fin is vibily joining highlighted in pink, resembling the caudal fins of \textit{Noturus}. This change is analogous to \Cref{fig:trait_swapping1}, where level-2 information of \textit{Noturus} was replaced with \textit{Notropis} (vice-versa), resulting in the caudal (tail) fin getting forked or split. Hence, this helps us understand that at Level-2, the two species diverged to develop different caudal fins. However, for PhyloNN, the generated image after trait-swapping is blurry, and most of the traits still closely resemble close the source species, which is unlikely given that the level-2 embeddings have been replaced.

\clearpage
\section{Additional Samples of Generated Images}\label{app:additional-samples}

\begin{figure}[ht]
\centering
\includegraphics[width=1\linewidth]{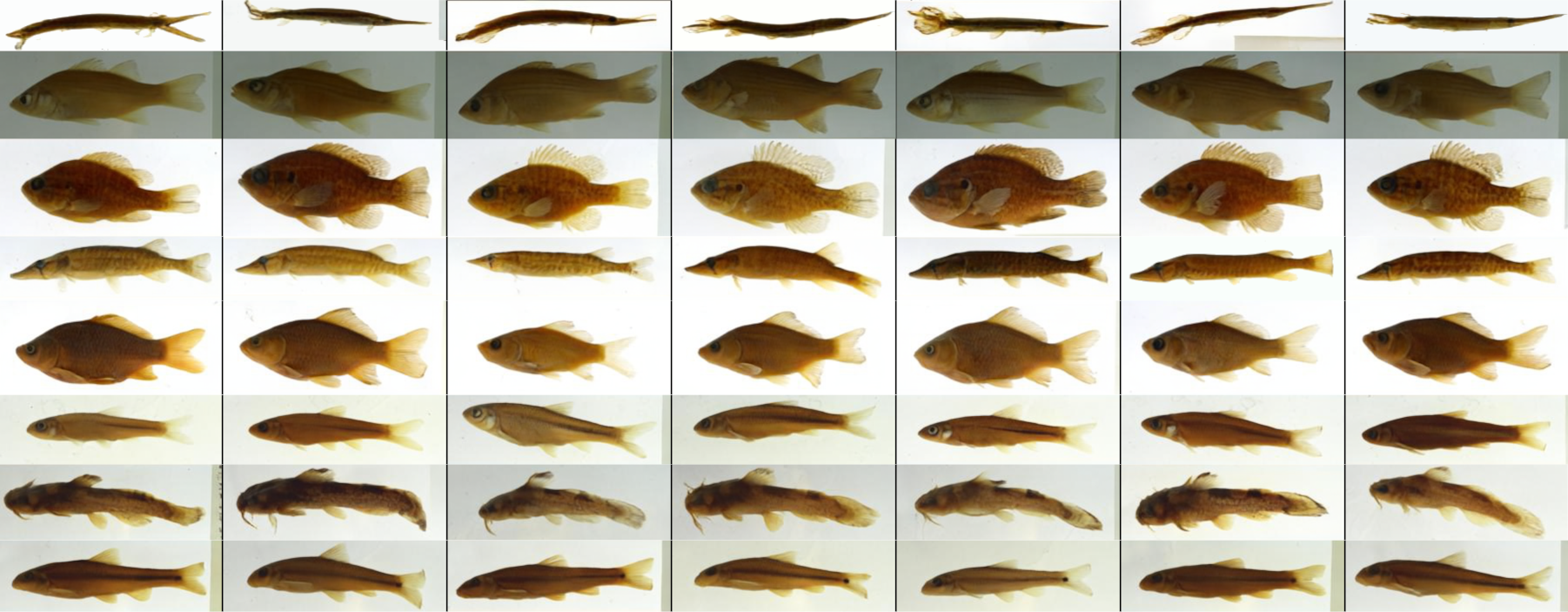}
\caption{Comparison of images of different species generated by HEIR-Embed where the generated images for a given row depict variations for the same species while the different rows represent distinct species. The order of species from top to bottom is \textit{Lepisosteus osseus, Morone chrysops, Lepomis gulosus, Esox americanus, Carassius auratus, Notropis blennius, Noturus exilis, Phenacobius mirabilis}}
\label{fig:hle_sample_grid}
\end{figure}

\Cref{fig:hle_sample_grid} shows additional examples of generated images for different species using Phylo-Diffusion. Each row of the figure depicts the generated images for the same species while the different rows represent distinct species. Notably, we observe inter-class variations among species belonging to the same class, such as differences in fish orientation and size.

\section{Ablation Results} \label{app:ablations}

\subsection{Generalization to Unseen Species: Leave-three-out}
As an additional ablation experiment, we conduct a leave-three-out experiment by excluding three species from different subtrees during training to test the model's ability to generalize to new species and situate them in the phylogeny. This experiment involves training the model excluding three species, \textit{Notropis blennius, Noturus gyrinus, and Lepomis humilis} that belong to different subtrees as seen in \cref{tab:groupings}. The generated images from the three subtrees after trait masking closely resemble the actual images of the 3 species, with an F1 score of 95.6 on a classifier trained to discriminate the 3 species. This experiment underscores the robustness and accuracy of Phylo-Diffusion in embedding and generating phylogenetically consistent images.

\clearpage

\subsection{Effect of Varying the Number of Levels in Phylo-Diffusion}
To demonstrate the robustness of Phylo-Diffusion, we perform ablation experiments with varying numbers of levels in the discretization of the phylogeny tree. We show that the choice of the number of levels depends on the depth of the phylogenetic tree and the internal nodes to be studied.
We train models with $\{2, 4, 6,8\}$ levels on the phylogeny shown in \Cref{fig:phylogeny}. \Cref{tab:ablations_num_levels} demonstrates that the model is robust to the choice of the number of levels.

\begin{table}
\centering
\caption{Quantitative results for Phylo-Diffusion with varying number of levels in the discretized phylogeny tree.}
\begin{tabular}{cccccc}
\hline
 \textbf{\# levels} & \textbf{FID ↓} & \textbf{IS ↑}  & \textbf{Prec. ↑} & \textbf{Recall ↑} \\ \hline

2 & 11.84 & 2.45 & 0.67 & 0.36 \\
4 & 11.38 & 2.53 & 0.65 & 0.37 \\
6 & 11.41 & 2.49 & 0.66 & 0.37 \\
8 & 11.77 & 2.50 & 0.67 & 0.37 \\ \hline
\end{tabular}
\label{tab:ablations_num_levels}
\end{table}

\subsection{Effect of Varying Embedding Dimensions in Phylo-Diffusion}
We further evaluate the effect of varying the number of embedding dimensions used in HIER-Embed on the performance of Phylo-Diffusion. In this experiment, we trained Phylo-Diffusion by varying HIER-Embed's dimension in the following range of values: $\{16,32,64,128,256,512,1024\}$. \Cref{tab:ablations_embed_dim} shows that Phylo-Diffusion is quite robust to the choice of embedding dimension with minimal drop in performance as we reduce the embedding dimension even to small values.

\begin{table}
\centering
\caption{Quantitative results for Phylo-Diffusion with varying embedding dimensions of hierarchical embeddings.}
\begin{tabular}{cccccc}
\hline
 \textbf{Embedding Dim.} & \textbf{FID ↓} & \textbf{IS ↑}  & \textbf{Prec. ↑} & \textbf{Recall ↑} \\ \hline

16 & 11.23 & 2.45 & 0.66 & 0.36 \\
32 & 11.25 & 2.45 & 0.66 & 0.38 \\
64 & 11.56 & 2.47 & 0.66 & 0.37 \\
128 & 11.31 & 2.45 & 0.67 & 0.37 \\
256 & 11.53 & 2.42 & 0.67 & 0.35 \\
512 & 11.38 & 2.53 & 0.65 & 0.37 \\
768 & 11.69 & 2.49 & 0.65 & 0.37 \\
1024 & 11.51 & 2.48 & 0.67 & 0.36 \\ \hline
\end{tabular}
\label{tab:ablations_embed_dim}
\end{table}

\clearpage
\section{CUB Dataset Results} \label{app:cub_results}
To show the applicability of our approach on other datasets with larger and deeper phylogenies, we perform additional experiments on 190 bird species from the CUB-200-2011 dataset (see \Cref{tab:cub_fid}). We selected the set of bird species based on whether we are able to obtain their phylogenetic knowledge from \href{https://birdtree.org/}{Bird Tree}, which are pre-processed similar to the fishes. We removed the background of these images using segmentation masks to focus only on the body of the birds.

\begin{table}
\centering
\caption{Quantitative results on the new birds dataset (30 samples/class). The classifier has a base accuracy of 76\% on the test set.}
\begin{tabular}{lccccc}
\hline
 \textbf{Method} & \textbf{FID ↓} & \textbf{IS ↑}  & \textbf{Prec. ↑} & \textbf{Recall ↑} & \textbf{F1 ↑} \\ \hline
Class Conditional & 6.8 & 3.2 & 0.70 & 0.49 & 0.68  \\
Scientefic Name & 8.5 & 3.1 & 0.65 & 0.48 & 0.18 \\
Phylo-Diffusion (ours) & 6.7 & 3.1 & 0.72 & 0.49 & 0.64 \\ \hline
\end{tabular}
\label{tab:cub_fid}
\end{table}

\subsubsection{Trait Masking:} 
\begin{figure}
\centering
\includegraphics[width=1\linewidth]{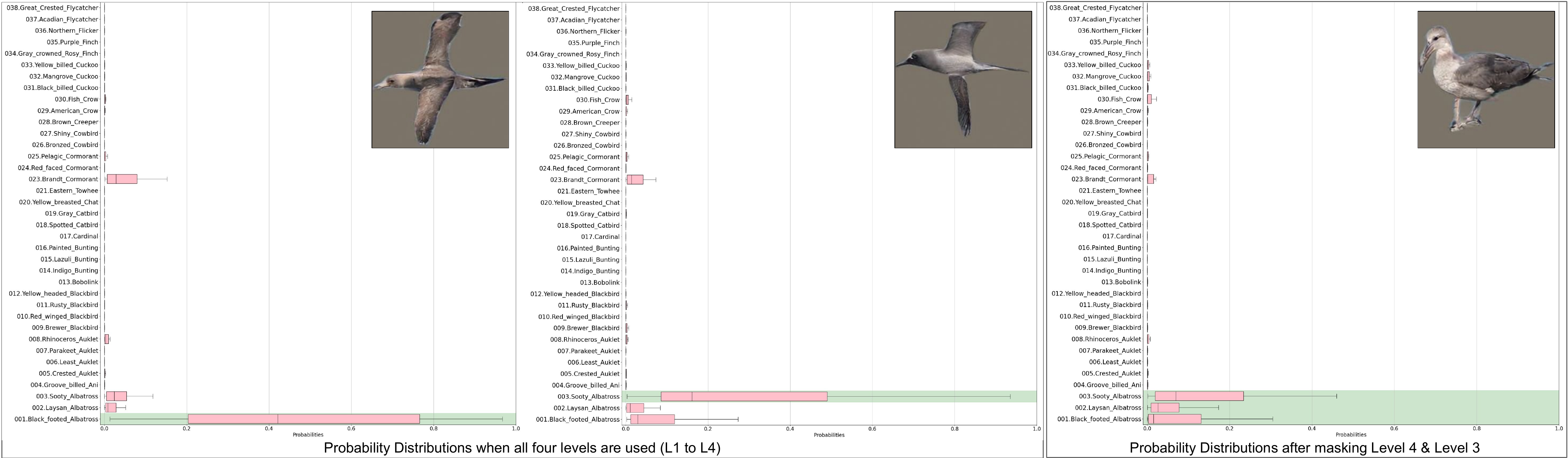}
\caption{Left: class probability distributions of images generated by using embeddings at all four levels for two species \textit{Black-footed albatross} and \textit{Sooty albatross} (shown in green) that are part of the same sub-tree till level 2. Right: class probability distributions of images generated by masking level 3 and level 4 (descendant species that have common ancestry till level 2 are highlighted in green).}
\label{fig:cub-level2-class0}
\end{figure}

Similar to fishes, \Cref{fig:cub-level2-class0} shows the changes in probability distributions when Level 3 \& 4 information is replaced with \textit{noise}. The first two plots show the logits of images generated for \textit{Black-footed albatross} and \textit{Sooty albatross} using embeddings from all the four levels. The third plot shows the dispersion of logits across the three descendant species that are part of the sub-tree defined till level 2, i.e., masking level 3 \& 4. We see similar results for CUB as well where the probability of classifying the generated images into any of the descendant species that share a subtree (highlighted in green) is generally greater than the species outside the subtree. 

\clearpage

\subsubsection{Trait Swapping:} 
\begin{figure}
\centering
\includegraphics[width=1\linewidth]{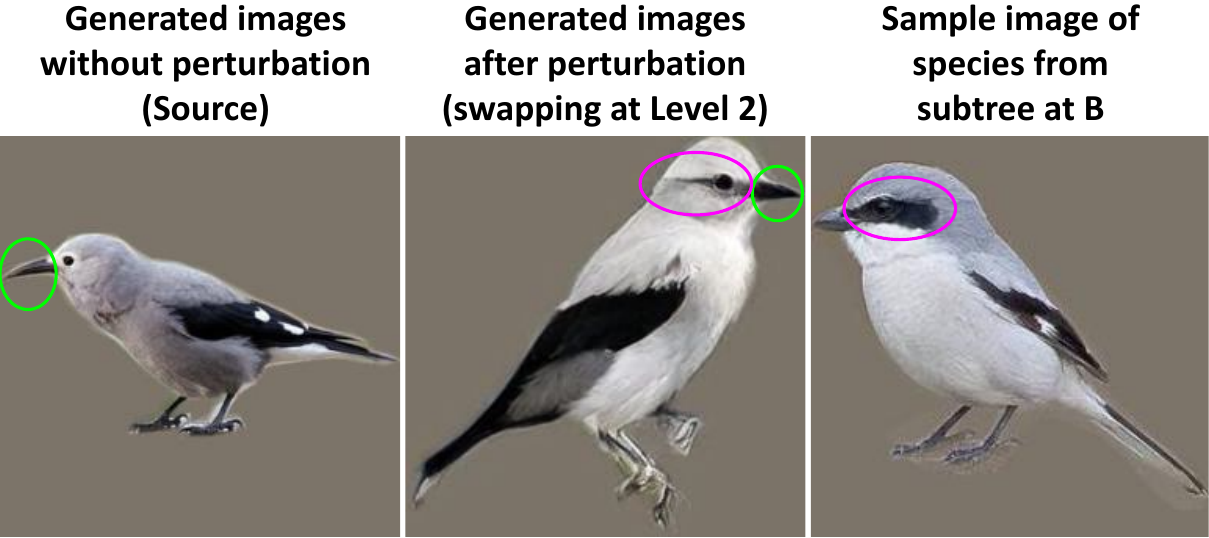}
\caption{Visualization of changes in traits after swapping information at Level 2 for \textit{Clark nutcracker} (left) with its species from its sibling subtree (right) to generate perturbed species (center). Traits shared with the source species are outlined in green, whereas those shared with the sibling subtree at Node B are outlined in pink.}
\label{fig:cub-trait_swapping}
\end{figure}

\Cref{fig:cub-trait_swapping} shows an example of the trait swapping experiment on the birds dataset, similar to the experiments for fishes in the main paper. We see that the image generated from the perturbed embedding (center) picks up the trait of black coloration around the eye (purple circle) that is shared by the target sub-tree (right) while traits like pointed beak (green circle) are retained from the source species (left).

\end{document}